\documentclass[natbib]{svjour3}                     
\smartqed  
\usepackage{graphicx}
 \usepackage{aps-bibstyle}  
\begin{document}

\title{The Solar-Stellar Connection}

\titlerunning{The Solar-Stellar Connection}        

\author{A.S. Brun,  R.A. Garc\'\i a, G. Houdek, \\ D. Nandy, M. Pinsonneault}


\institute{Brun, A.S.  and  Garcia, R.A. \at
	      AIM, CEA/CNRS/University of Paris 7, CEA-Saclay, 91191 Gif-sur-Yvette, France\\
               \email{sacha.brun@cea.fr and  rafael.garcia@cea.fr}
               \and
               Houdek, G.
               \at Stellar Astrophysics Centre (SAC), Department of Physics and Astronomy, Aarhus University, Ny Munkegade 120, DK-8000 Aarhus C, Denmark \\
               \email{hg@phys.au.dk}
               \and
               Nandy, D.
               \at Center of Excellence in Space Sciences India (CESSI) and Department of Physical Sciences, Indian Institute for Science Education and Research Kolkata, Mohanpur 741252, West Bengal, India\\
               \email{dnandi@iiserkol.ac.in}
               \and
               Pinsonneault, M.
               \at Department of Astronomy, The Ohio State University, 140 West 18th Avenue, Columbus, OH 43210, USA \\
               \email{pinsonneault.1@osu.edu}
}

\date{Received: date / Accepted: date}

\maketitle

\begin{abstract}
We discuss how recent advances in observations, theory and numerical simulations have allowed the stellar community to progress in its understanding of stellar convection, rotation and magnetism and to assess the degree to which the Sun and other stars share similar dynamical properties. Ensemble asteroseismology has become a reality with the advent of large time domain studies, especially from space missions.  This new capability has provided improved constraints on
stellar rotation and activity, over and above that obtained via traditional techniques such as spectropolarimetry or CaII H\&K observations. New data and surveys covering large mass and age ranges have provided a wide parameter space to confront theories of stellar magnetism. These new empirical databases are complemented by theoretical advances and improved multi-D simulations of stellar dynamos.  We trace these pathways through which a lucid and more detailed picture of magnetohydrodynamics of solar-like stars is beginning to emerge and discuss future prospects.

\end{abstract}

\section{Solar Dynamics and Magnetism in a Stellar Context}
\label{intro}

Time-dependent stellar magnetic fields, such as that of the Sun, are generated in their interior through a magnetohydrodynamic (MHD) dynamo mechanism that involves interactions between plasma flows and magnetic fields \citep[e.g.][]{Moffatt:1978tc,charbonneau10}. The generated magnetic fields are manifested at the surface through starspots or polar caps, and contribute to the dynamic activity of the star. This activity governs the structure of stellar coronae, the stellar radiative flux output, and transient but energetic events such as flares \citep{gudel04,2012JGRA..117.8103S}.  Stellar activity can also have a strong impact on the properties of stellar astrospheres which may host planets \citep{2004LRSP....1....2W}. Magnetized solar-like winds extract angular momentum which drives angular momentum in low mass stars. Understanding the origin of stellar magnetism is therefore of crucial importance, and remains an outstanding problem in astrophysics.

The Sun, our nearest star, provides easy access to spatially resolved observations of magnetic activity and thus sets constraints on the dynamo origin of stellar magnetic fields. However, significant uncertainties in our current understanding of the solar dynamo mechanism motivates us to confront our theoretical ideas of magnetic field generation with the wider space parameters available from stellar activity observations \citep[][and references therein]{noyes84b,baliunas95,wright04,2005ssac.conf..307G,2007ApJ...657..486B,2009A&A...501..703O,2009ARA&A..47..333D,2012LRSP....9....1R}. While, in principle, different stars with different properties may host distinct classes of MHD dynamos, a good starting assumption is that a reasonable model of solar and stellar magnetism should be able to reproduce the range of stellar activity taking into account appropriate changes in the internal driving mechanisms. For example, high mass stars have shallow surface convective zones with high turbulent fluxes relative to low mass stars. Stars with different rotation rates and ages will have different magneto-convective properties (usually characterized by varying non-dimensional parameters such as the Reynolds or Rossby numbers). To reduce this problem to a practical exercise, it is also useful to deal with solar-like stars (narrowing the range of spectral type considered), within which, there is substantial variation in the nature of magnetic output \citep{2008MNRAS.388...80P,2013arXiv1311.3374M}. The observation and characterization of stellar convection, rotation and magnetism has undergone a recent boom with the outcome of better ground-based instruments and space-born missions. For instance, until recently chromospheric proxies such as emission in the core of {\rm CaII H \& K} lines and white light modulation were the most commonly used techniques for characterizing stellar magnetism \citep{wright04,2005ssac.conf..307G}. The development of very sensitive spectropolarimeter such as ESPaDOnS or NARVAL has
helped to further improve our understanding of stellar magnetism by providing detailed magnetic maps and sometimes the surface differential rotation of the star \citep{1997MNRAS.291..658D,1999MNRAS.305L..35J,2006MNRAS.370..629D}. With the success of planet transit/asteroseismology missions such as Corot, \emph{Kepler}, the picture can now be refined even further \citep{2014arXiv1403.7155G}.
We have now entered an era in which systematic coupled analyses of chromospheric activity proxies, spectropolarimetry maps and asteroseismic data is being performed.  This allows us to better characterize magnetism of solar-like stars and its variability on short and relatively long terms \citep{berdyugina05,2008MNRAS.388...80P,2011AN....332..866M,2011sf2a.conf..497M,2013AN....334...48M,2013A&A...550A..32M}. The purpose of this review paper is to recap recent progress made, with the aid of asteroseismology, theory and numerical simulations, in our understanding of stellar dynamics and magnetism in solar-like stars. Other proxies, such as the time evolution of surface lithium abundances, can also be useful to constrain
stellar dynamics but they are outside the scope of this review \citep[for discussions see][and references therein]{1999ApJ...525.1032B,2004A&A...418.1051T,2010A&A...522A..26S,2010A&A...519A.101D,2012MSAIS..22....9S}.
We begin by discussing recent observational findings with a special emphasis on ensemble asteroseismology and on stellar dynamics, magnetism and cyclic activity.
We then present the current theories of stellar rotational history based on 1-D models and of stellar magnetism with an introduction to dynamo theory based on 2-D kinematic dynamo models and 3-D nonlinear turbulent numerical simulations of stellar dynamo.
We continue with a discussion on how theoretical models hold up against observations such as scaling laws and rotation-activity relationship \citep{2003A&A...397..147P,2011ApJ...743...48W} and provide a critical assessment of the current understanding. Finally we conclude by proposing possible future developments in asteroseismology and stellar dynamics to better constrain solar-like star rotation and activity.

\subsection{New Asteroseismic Diagnostics of Global Stellar Convection, Rotation and Magnetism}

Oscillations are intrinsically damped and stochastically excited by the turbulent motion in the outer convective layers of cool stars \citep[e.g.][]{1977ApJ...212..243G,1992MNRAS.255..603B,1999A&A...351..582H,2005MNRAS.360..859C,2008A&A...478..163B,2010Ap&SS.328..237H}. We refer to stars exhibiting such an oscillation pattern as solar-type stars hereafter.  These oscillations are also present with higher amplitudes in red-giant stars \citep[e.g.][]{2002MNRAS.336L..65H,2009Natur.459..398D,2010ApJ...713L.176B,2010A&A...517A..22M} (and even in mores massive B-type stars as recently shown by \citealt{2009Sci...324.1540B}). The frequency of maximum power scales as $(GM/R^2) T_{\rm eff}^{-1/2}$ \citep[e.g.,][]{1995A&A...293...87K}; given the wide dynamic range in stellar surface gravities the relevant periods span the range of minutes to weeks.

The magnetic activity cycles in the Sun and other solar-like stars are often the consequence of dynamo processes induced by the interaction of three fundamental mechanisms: convection, rotation, and magnetic fields (see detailed discussion in \S 2).  Asteroseismic observations place powerful observational constraints on each of these three processes as we will see in \S 1.4, 1.5, and 1.6. For example, activity-cycle induced variations of pulsation properties (mode linewidth and amplitudes) in the Sun and in solar-like stars were discussed, for example, by \cite{2001MNRAS.327..483H} and \cite{2007MNRAS.377...17C}.

The long and uninterrupted high-precision photometric observations required for asteroseismology are ideal for characterizing long-term variations such as the ones induced by starspots crossing the visible disk of stars. Indeed, the dimming produced by these spots leads to a measurable modulation in the photometry with a period corresponding to the surface rotation at the active latitudes \citep[e.g.][]{2011AJ....141...20B,2012ASPC..462..133G,2014arXiv1403.7155G,2013ASSP...31..215L,2013MNRAS.436.1883W,2013A&A...557L..10N,2013arXiv1308.1508R,2014ApJS..211...24M} in which the spots develop. One must keep in mind that on the Sun faculae and the small-scale field around over-compensate this dimming effect, leading to a net brightening at solar maximum \citep{2000SSRv...94..113S,2012SGeo...33..453F}. Moreover their relative amplitude depends on how close to the limb the starspot is and to the stellar inclination angle with respect to the line of sight. Mapping the relative ratio of dimming through star spot vs. brightening through faculae and small-scale field changes as a function of stellar spectral type, age, rotation rate and metallicity is an emerging field, and as of this paper the overall pattern is not firmly established. Some studies have started to address the physical origins of stellar variability \citep[e.g.][]{2014arXiv1406.2383S} and found that a transition from net brightening to net dimming should occur around an index of chromospheric activity $\log(R'_{\rm HK})$ of -4.9, a value which is consistent with observations \citep{2007ApJS..171..260L,2009AJ....138..312H}.

Ground based stellar variability studies have a long history, and they have traditionally focused on young and active stars.  The Convection Rotation and planetary Transits (CoRoT) and \emph{Kepler} space missions have contributed continuous and precise data series for large samples, which has had a dramatic impact on the quantity and quality of stellar rotation data.  CoROT and \emph{Kepler} have extended our understanding to large numbers of older and less active field stars \citep[e.g.][]{2009A&A...506..245M,2010A&A...518A..53M,2011ApJ...733...95M,2011A&A...534A...6C}. Asteroseismology has also extended our understanding of longer term cycles.  In the solar case the number of spots in the Sun changes during the solar magnetic cycle \citep{2010LRSP....7....1H}, introducing a modulation of the variance in the photometric observations that can be used as a proxy of the magnetic activity cycle in the Sun \citep{2009ASPC..416..529M} and in other solar-like stars \citep[e.g.][]{2010Sci...329.1032G,2011ApJ...732L...5C}.  As the temperature rises from the photosphere to the chromosphere, the cores of very strong lines, such as {\rm Ca II H} (396.8 nm) and {\rm K} (393.4 nm) emission lines \citep{1959ApJ...130..366L,1978ApJ...226..379W,baliunas95}, develop emission features whose strength depends on the magnitude of the temperature inversion. However, seismic techniques can go further by probing structural and dynamical changes beneath the photosphere as the magnetic cycles develop. Indeed, since the dawn of helioseismology, several authors noticed that the frequency of the acoustic modes increase near the solar maximum \citep{1985Natur.318..449W,1987A&A...177L..47F,1989A&A...224..253P,1990Natur.345..779L,1992A&A...255..363A}. Nowadays, it is possible to measure variations in all the p-mode properties not only in the Sun  \citep[e.g.][]{2000MNRAS.313...32C,JimCha2007}, but also in other stars such as the F star HD49933 measured by CoRoT, in which the amplitude as well as the frequency shifts of low-degree modes change with time \citep{2010Sci...329.1032G}.

Asteroseismic measurements can also be used to extract information about convection. The p modes are observed as a power excess at distinct frequencies on top of a continuous Fourier spectrum. This background spectrum is likely due to the presence of convective motions at all scales with granulation playing a key role \citep[e.g.][]{harvey85,2008A&A...490.1143L}, but may also be due to the presence of faculae \citep[e.g.][]{2013arXiv1302.5563K}. Until now, neither classical observations nor one-dimensional analytical convection models could put constraints on granulation. The situation has changed with asteroseismic observations of thousands of stars thanks to the CoRoT \citep{2006cosp...36.3749B} and \emph{Kepler} \citep{2010Sci...327..977B} missions. For instance, \citet{2011ApJ...741..119M} have determined the timescale and power of granulation for around 1000 \emph{Kepler} red giants at different stages of their evolution going from the red giant branch up to the clump. Moreover, the depth $d_{\rm bcz}$ of the convective zone  --which is often the location used to compute a caracteristic convection overturning time e.g.
$\tau_{\rm conv} \sim d_{\rm bcz}/v_{\rm conv}$, with $v_{\rm conv}$ being a characteristic convective velocity --  is a region characterized by a sharp change in stratification, since it changes from (super)-adiabatic to subadiabatic, modulo the extent of convective penetration \citep[e.g.,][]{1991A&A...252..179Z}. Therefore, the sound speed undergoes an abrupt variation producing an acoustic glitch that leaves a fingerprint in the frequencies of the acoustic modes \citep[e.g.,][]{1988IAUS..123..155G,1988IAUS..123..151V,1990LNP...367..283G} see also \citep{1994A&A...283..247M,1995MNRAS.276.1402B,2007MNRAS.375..861H,2011MNRAS.414.1158C,2014ApJ...782...18M}.

\subsection{The Link Between Helio- and Asteroseismology: Global Oscillations and 1-D Models}

Helioseismology has been providing a wealth of high-quality information about
the structure and physics of the solar interior since the first
detection of the global five-minute oscillation modes \citep{1979Natur.282..591C}.
The large number of observable global solar oscillation modes,
with spherical degrees up to several thousands, has allowed
helioseismologists to infer with high accuracy from the measured oscillation
frequency the solar sound speed and density
as a function of solar radius. Such helioseismic inversion techniques make use of the fact
that the displacement eigenfunctions are related to the
adiabatic eigenfrequencies through a variational principle
\citep{1963ApJ...138..896C,1967MNRAS.136..293L} leading
to a linear relation between a small perturbation in the
density (and squared sound speed) of a solar model and
the corresponding change in the frequencies \citep[e.g.][]{1978Natur.274..739G,Gough1984}.
Before the results of the Sudbury Neutrino Observatory (SNO) experiment (see further down),
the solar neutrino problem, together with the high-quality helioseismic data collected from
space, indicated clear problems in our understanding of either basic physics or the structure of the solar interior. In response, there was a major sustained effort to improve
solar models; for a review see e.g. \cite{2004SoPh..220..137C}
and references therein.
The rather good agreement ($<1\%$) between the Sun and 1-D solar models
computed in the late 1990s \citep{1996Sci...272.1286C,1998ApJ...506..913B,1999ApJ...525.1032B} was, however, degraded if the solar
photospheric abundances reported by \cite{2004A&A...417..751A} were adopted. The problem still holds even after the revision to
a slightly higher value of metalicity by \cite{2009ARA&A..47..481A}. We refer to \cite{2010Ap&SS.328...51C}, \cite{2014ApJ...787...13V} and chapter 2 \citep[e.g.,][]{2014SSRv..tmp....3B} of the ISSI book \ref{sismoissi} for a recent discussion on helioseismology constraints on metallicity/abundances.

Still, the high-quality helioseismic data has led to many fundamental achievements in solar
physics such as inverting the internal solar angular-velocity profile or helping to disentangle the long-standing solar
neutrino problem. Seismic inversions have revealed that the non-uniform, surface
differential rotation extends all the way to the base of the convective envelope and becomes almost
uniform (e.g. solid body rotation) in the radiative interior
\citep[e.g.][]{1987ApJ...314L..21B,1990SoPh..125....1T,ThoJCD2003,2004SoPh..220..269G,2008A&A...484..517M,2012SoPh..tmp..149E}.  The inferred radial interior density and sound-speed profiles also place strong constraints on the standard
solar model and consequently on the theory of stellar structure and evolution. Since the neutrino fluxes predicted by
solar models could not be explained by standard zero mass neutrino physics, particle physicists had
to consider massive neutrinos and their associated flavor oscillations \citep[e.g.][]{STCCou2001}.
Thanks to the SNO experiment measurements \citep{2001PhRvL..87g1301A} the scientific community has
since confirmed, by measuring both charged and neutral currents, that the original deficit of electronic
neutrinos is due to their transition to another neutrino flavor as
they travel from the solar core to Earth's detectors.
Another important issue, which has again become topical very recently by analyzing the high-quality data from the
Heliospheric and Magnetic Imager (HMI) on NASA's Solar Dynamics Observatory
(SDO), is the solar oblateness \citep{2012Sci...337.1638K} and the consequent
distortion of the Sun's gravitational field which may account, in part, for
the remaining 0.2\% discrepancy in the precession of Mercury's orbit
\citep[e.g. ][]{1998ApJ...505..390S,2012Sci...337.1611G}. Detailed discussions about what we
have learnt from helioseismology and how helioseismology can help to address
the remaining problems in solar physics can be found in recent reviews by
\cite{2004SoPh..220..137C} and \cite{2013SoPh..287....9G} and all the chapters of the ISSI book \ref{sismoissi}.

That stellar variability provides
deep insights into stellar structure has been known since the
beginning of the 20th century \citep[e.g.][]{1914ApJ....40..448S}. In nonradial oscillators such as the Sun, the numerous
pulsation modes with distinct frequencies encode detailed information about their interior structure. The Sun
oscillates in approximately 10 million distinct oscillation modes and
with the help of space-borne instruments we are now able to
measure some 6000 individual modes and up to several million modes via so-called ridge fitting.
The amplitudes of solar-like oscillations, however, are small
compared to those observed in classical pulsators, such as
Cepheids or Delta Scuti stars, and their frequencies can also be quite high (5 minutes in the solar case). Therefore only the latest generation
of instruments have been able to detect solar-like
oscillations in stars other than the Sun.

The first indication of excess acoustic power of solar-like oscillations in
other stars with a frequency dependence similar to the Sun was reported
by \citet{1991ApJ...368..599B} for the F5 star Procyon A ($\alpha$ CMi), for which
the first unambiguous detection of solar-like oscillations
was reported by \citet{1999A&A...351..993M} from the ground, as well as by \citet{2001ESASP.464..391S}
using data collected by the WIRE spacecraft. These results were confirmed later
by \citet{2008A&A...478..197M}. The first (unconfirmed) detection of individual peaks
in the acoustic power spectrum from high-precision time-resolved spectroscopic
observations was published for the G0 star $\eta$ Boo by \citet{1995AJ....109.1313K}, but it
was not until 2003 that an unambiguous confirmation was established by \citep{2003aahd.conf..315C,2003AJ....126.1483K}.

High-degree oscillation modes will not be accessible in other stars
in the foreseeable future because they require spatially resolved variability data.  However, the observable low-degree modes in distant stars
still provide rich information about the internal structure and dynamics of a large number
of solar-like stars at different evolutionary stages, as demonstrated recently
by the results obtained by the CoRoT and \emph{Kepler}  missions.

Additional projects, such as the Sun-in-time \citep{2007LRSP....4....3G}
or the Sun as a star \citep{2005ssac.conf..307G}, try to characterize the Sun
by comparing its property to solar-like stars at various evolutionary stages.
In particular, the understanding of solar rotation, convection, variability
and magnetic activity in a stellar context allows to better constraint theory by deriving key scaling laws
as a function of age, rotation, chemical abundances and mass of
those various physical processes. We defer to \S 1.4 to 1.6 for discussions of the relevant observations
and to \S 2 for the description of the current theoretical understanding.

Returning to asteroseismic diagnostic, even with the limited seismic information of low-degree oscillation modes, some theoretically
important properties have become accessible, such as the gross
structure of the energy-generating core and the extent to which it is
convective, and a large-scale core to envelope contrast in the angular velocity of evolved stars
\citep[e.g.][and references therein]{2012A&A...540A.143M,2012ApJ...756...19D,2012ApJ...749..152M,2013ASSP...31..171G}.
Such information will be of crucial importance for checking, and then calibrating, the theory of
the structure and evolution of stars.

One of the many interesting applications of utilizing low-degree modes for
measuring various aspects of the structure in distance stars is the
measurement of rapid variations in the stellar background, brought about by,
for example, the rapid variation in the acoustic cutoff frequency at the
base of the surface convection zone or the variation in the first adiabatic
exponent in the helium ionization zones. The design and analysis
of low-degree seismic signatures of the structure of spherical stars is
therefore of central importance to asteroseismic diagnosis, and various
research groups have been working on improved seismic diagnostic
techniques \citep[e.g.][and references therein]{2007MNRAS.375..861H}. First
applications of low-degree seismic diagnostic techniques have been carried out
recently on 19 solar-type stars observed by the \emph{Kepler} mission \citep{2014ApJ...782...18M}.

However our poor understanding of the physics of the stellar surface layers causes
a major problem in the analysis of global solar-like oscillation, because the properties of the near-surface layers have a major impact on oscillation frequencies.
Typical stellar structure calculations treat the superficial layers with
simplified atmosphere models and the stratification of the superadiabatic
region by means of a local, time-independent mixing-length approach
\citep[e.g.][]{1958ZA.....46..108B}. The effects of the (turbulent) Reynolds stresses
in both the hydrostatic equilibrium and pulsation calculations are essentially
always ignored. Linear pulsation calculations typically adopt the adiabatic
approximation and ignore the momentum flux perturbations (turbulent pressure
fluctuations). However, nonadiabatic effects and the fluctuations of the
turbulent fluxes (heat and momentum) do modify the modelled pulsation
eigenfunctions and consequently also the oscillation frequencies
\citep[e.g.][]{1984AdSpR...4...85G,1992MNRAS.255..603B,1996PhDT........80H,2010Ap&SS.328..237H}.
For a ``standard'' solar model, such as Model S
\citep{1996Sci...272.1286C}, the frequency differences, scaled with
the mode inertia, between the solar and model frequencies are dominated
by the near-surface effects and are predominantly a function of frequency
alone. Differences between solar frequencies observed with the GONG instruments
and adiabatically computed Model S frequencies are illustrated by the
the symbols (plusses) in the left panel of Figure\,1 (colours indicate
different spherical degrees). The increase of the frequency residuals with
oscillation frequency depends on the modelling details of the functional
form of the acoustic cutoff frequency (as it affects the acoustic
potential) with radius in the near-surface layers. In the Sun the acoustic
cutoff frequency is about 5.5 mHz.

Sophisticated 3-D hydrodynamical simulations of stellar convection have
been used to estimate the effect of turbulent pressure in the equilibrium model (the mean Reynolds stress).
For example, \cite{1995ESASP.376b.459R} investigated the effect on adiabatic
eigenfrequencies of the contribution that the turbulent pressure makes to the
mean hydrostatic stratification. They examined a hydrodynamical
simulation by \cite{1991LNP...388..195S} of the outer 2\% by radius of the Sun,
matched continuously in sound speed to a model envelope calculated, as in
a ``standard'' solar model, with a local mixing-length formulation. The
resulting frequency shifts of adiabatic oscillations between the simulations
and the ``standard'' solar reference model, Model S, are illustrated in the
right panel of Figure\,1. The frequency residuals behave similarly to the
data (plus symbols) in the left panel of Figure\,1 though with larger
shifts at higher oscillation frequencies.

\begin{figure}
\begin{center}
\includegraphics[width=0.85\textwidth]{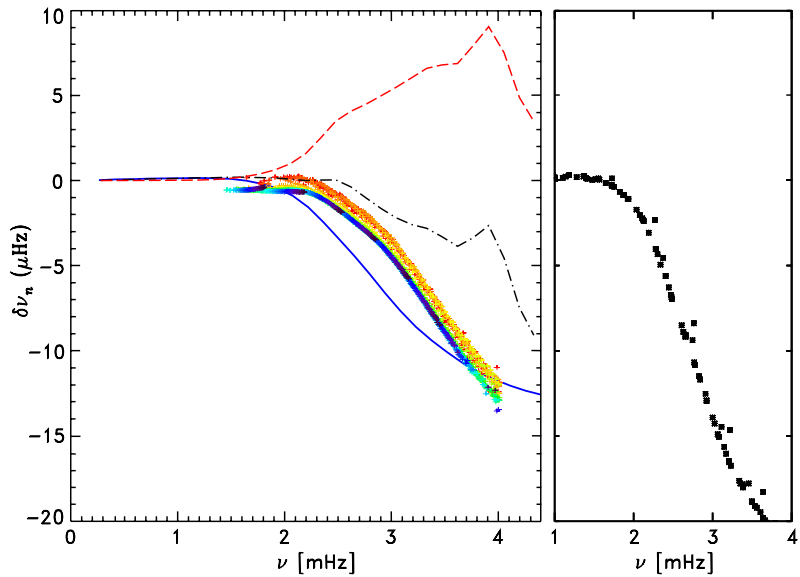}
\end{center}
\caption{{\bf Figure 1}: Frequency residuals between the Sun and solar
models as a function of frequency.
Left: scaled differences between observed GONG frequencies and adiabatically
computed frequencies of the ``standard'' solar Model S. Different colours
indicate different degrees of the oscillations modes
\citep[adapted from][]{1996Sci...272.1286C}. The curves show
frequency residuals between models computed with a nonlocal time-dependent
and a ``standard'' mixing-length formulation for convection: The solid (blue) curve
shows the adiabatic residuals cause by the Reynolds stress in the mean model. The dashed (red) curve is
the frequency shift caused by nonadiabaticity and convection dynamics. The
overall frequency shift is plotted by the dot-dashed curve \citep[adapted from][]{1996PhDT........80H}.
Right: Scaled adiabatic frequency differences between a model for which
the near-surface layers were represented by a hydrodynamical simulation,
and the ``standard'' solar Model S \citep[adapted from][]{1995ESASP.376b.459R}.}
\end{figure}

The effects of nonadiabaticity and convection dynamics on the pulsation
frequencies were, for example, studied by \cite{1992MNRAS.255..603B},
\cite{1995ESASP.376b.459R} and \cite{1996PhDT........80H}. In these studies the nonlocal,
time-dependent generalization of the mixing-length formulation by
\citet{1977ApJ...214..196G,1977LNP....71...15G} was adopted to model the heat and momentum flux consistently
in both the equilibrium envelope model and in the linear nonadiabatic stability
analysis. The outcome of such calculations is shown in the
left panel of Figure\,1 \citep{1996PhDT........80H,2010Ap&SS.328..237H} by the dot-dashed curve.
Similar as for the hydrodynamical simulations the effect of the Reynolds
stresses in the mean structure decreases the adiabatic
frequencies (solid curve) with a maximum deficit of about
12$\mu$Hz, which is much smaller than the deficit of about 24$\mu$Hz
from the hydrodynamical simulations (symbols in the right panel).\\
The effects
of nonadiabaticity and convection dynamics (dashed curve in the left panel),
however, lead to an increase of the mode frequencies by as much as
$\sim$9$\mu$Hz, nearly cancelling the downshifts from the effect of the
Reynolds stress (turbulent pressure) in the mean structure (solid curve
in the left panel), leading to the overall effect indicated by the
dot-dashed  curve in the left panel. If the positive frequency shifts between
models computed with and without nonadiabatic effects and convection dynamics
(dashed curve in the left panel) are interpreted as the nonadiabatic and
momentum flux corrections to the
oscillation frequencies then their effects are to bring the frequency residuals
of the hydrodynamical simulations (right panel) in better
agreement with the data (plus symbols in the left panel of Figure\,1).
If we were to understand the reason why the semi-analytical, nonlocal
convection model underestimates the effect of the Reynolds stress in the
mean model on the adiabatic eigenfrequencies, we would come closer to a
self-consistent explanation for the still unsolved problem of reproducing
theoretically the observed high-frequency oscillations in the Sun and in
sun-like stars.
Currently only simple procedures for estimating the near-surface frequency
corrections \citep{2008ApJ...683L.175K} are adopted by the community
\citep[e.g.][]{2012ApJ...749..152M}. This simple
procedure is based on the calculations of a model, in which the frequency
shifts are scaled by a power-law in frequency \citep{1980Natur.288..544C}.
A physically more realistic description of the surface effects with reliable
applications to solar-like stars is therefore of imminent importance
for the whole community to determine more reliable radii and ages in distant
stars from the observed mean large frequency separation and small
frequency separation, for they depend crucially on the near-surface
frequency corrections \citep[e.g.][]{1980Natur.288..544C,2010Ap&SS.328..237H}.

\subsection{Asteroseismic Scaling Relations}
\label{sect2}

With the advent of space missions, asteroseismologists studying solar-like stars started to apply techniques originally developed for Sun-as-a-star observations to other stars. With only one or two solar-like stars per target field observed by CoRoT \citep[e.g. HD~49933, HD~181420 and HD~181906,][respectively]{2008A&A...488..705A,2009A&A...506...51B,2009A&A...506...41G}, it was possible to do a full analysis of the power spectrum, determine the individual p-mode frequencies and, after combining the seismic observables with the spectroscopic ones \citep[e.g.][]{2009A&A...506..235B}, perform a full modeling of each target using stellar evolution codes \citep[e.g.][]{2013A&A...549A..12M}. Later on, with the study of red giants in the CoRoT exoplanet field and the arrival of \emph{Kepler}, asteroseismologists faced a new challenge: the need to analyze massive samples of thousands of stars with high-quality light curves. The combination of the high data volume and the surface effects above made manual analysis impractical.  As a result, seismologists turned to characterizing the pulsation frequency patterns with a measurement of the typical frequency of oscillation and the typical frequency spacing between carefully selected and classified modes.  To do this for large sample, automatic pipelines were developed in order to extract these mean seismic observables: $\Delta{\nu}$ and $\nu_{max}$ \citep[e.g.][]{2009CoAst.160...74H,2009A&A...508..877M,2010MNRAS.402.2049H,2010A&A...511A..46M}.

The large frequency spacing, $\Delta \nu_{\ell} (n)= \nu_{n,\ell}-\nu_{n-1, \ell}$, is proportional to the sound travel time in the cavity in which the mode propagates and, therefore, it is proportional to the density inside the star: $\Delta \nu \approx (M/R^3)^{1/2}$. This relation shows a very good agreement with modeling \citep[e.g.][]{1986ApJ...306L..37U} and can be scaled from the solar values as \citep{1995A&A...293...87K}:
\begin{equation}
\label{dnuMR}
   \Delta \nu \approx \Delta \nu_\odot  \left( \frac{M}{M_\odot} \right)^{1/2} \left( \frac{R}{R_\odot}\right)^{-3/2} \;\; ,
\end{equation}
in which $\Delta \nu_\odot$=135.1 $\pm$ 0.1 $\mu$Hz, as derived by \citet{2011ApJ...743..143H} using 111 VIRGO subseries of 30-day, each spanning from 1996 to 2005 and analyzed  in the same way as the \emph{Kepler} data. More recent work \citep{2012ApJ...751L..36W,2013A&A...550A.126M} based on numerical simulations suggests the need for small correction terms.

In 1991, \citeauthor{1991ApJ...368..599B} suggested that the frequency at the maximum power of the p-mode bump, $\nu_{\rm{max}}$, should scale with the acoustic cutoff frequency \citep[see also][]{2011A&A...530A.142B}, and following \citet{1995A&A...293...87K} it could be scaled from the solar values according to:

\begin{equation}
\label{numaxMRT}
   \nu_{\rm{max}} \approx  \nu_{\rm{max}, \odot}   \left(\frac{M}{M_\odot}\right) \left(\frac{R}{R_\odot}\right)^{-2} \left(\frac{T_{\rm{eff}}}{T_{\rm{eff},\odot}}\right)^{-1/2} \;\; ,
\end{equation}
in which $T_{\rm{eff},\odot}$ = 5770 K, and $\nu_{\rm{max}, \odot}$ = 3090 $\pm$ 30 $\mu$Hz \citep{2011ApJ...743..143H}. This constraint is less strongly
motivated by theory and requires empirical guidance.

Therefore, from equations~(\ref{dnuMR}) and (\ref{numaxMRT}), one can obtain the stellar mass and radius as a function of the seismic variables $\nu_{\rm{max}}$ and $\Delta \nu$ --  assuming that $T_{\rm{eff}}$ is known independently of evolutionary models \citep[e.g.][]{2009A&A...503L..21M}:

\begin{equation}
{R} \approx {R_\odot} \left( \frac{\Delta\nu_\odot}{\Delta \nu}\right)^2 \left(\frac{\nu_{\rm max}}{\nu_{\rm{max},\odot}} \right) \left(\frac{T_{\rm eff}}{T_{\rm{eff,\odot}}} \right)^{1/2} \;\; ,
\end{equation}

\begin{equation}
M \approx M_\odot \left(\frac{\Delta\nu_\odot}{\Delta \nu} \right)^4 \left(\frac{\nu_{\rm max}}{\nu_{\rm{max},\odot}} \right)^3 \left(\frac{T_{\rm eff}}{T_{\rm{eff,\odot}}} \right)^{3/2} \;\; .
\end{equation}

\begin{figure}[!htb]
\begin{center}
\includegraphics[width=0.85\textwidth]{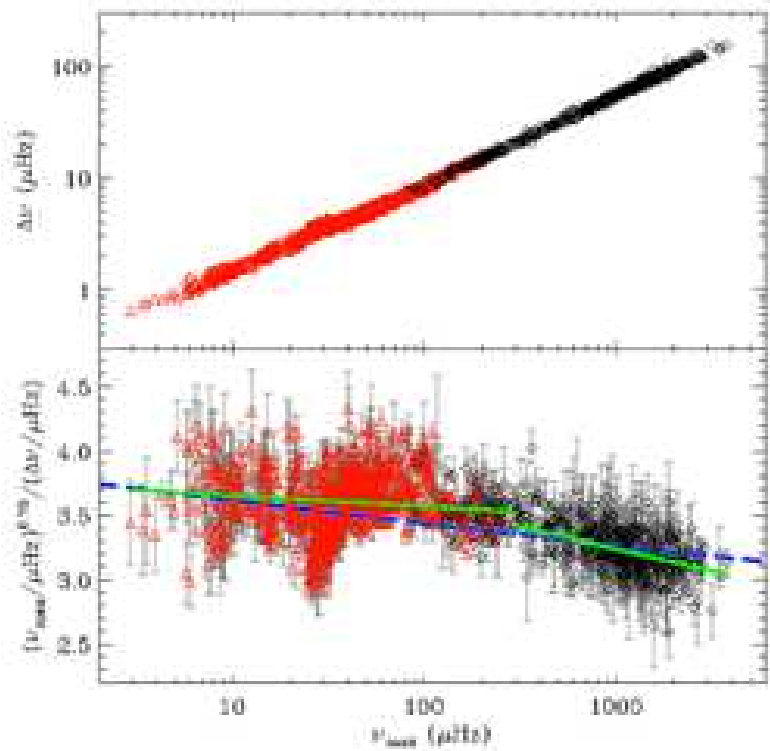}
\end{center}
\caption[$\Delta\nu$ versus $\nu_{\rm{max}}$ for 1700 stars observed by \emph{Kepler}]{\label{nmaxdnu} $\Delta\nu$ versus $\nu_{\rm{max}}$ (top panel) for 1700 stars observed by \emph{Kepler} taken from \citet{2011ApJ...743..143H}. The red triangles show stars observed in long cadence (red giants), while black diamonds are stars observed in short cadence (solar-like stars). In the bottom panel we have shown the same relation but after removing the luminosity dependence by raising $\nu_{\rm{max}}$ to the power of 0.75. Green lines show power law fits to the $\Delta\nu - \nu_{\rm{max}}$ relation for two different intervals of $\nu_{\rm{max}}$. The blue dashed line shows the relation of Eq.~\ref{dnu_numax} derived using both red giants and main-sequence stars \citep{2009MNRAS.400L..80S}.}
\end{figure}

It has also recently been shown \citep[e.g.][]{2009MNRAS.400L..80S,2009A&A...506..465H,2010A&A...517A..22M,2011A&A...525A.131H} that solar-like oscillations in main-sequence stars follow to a good approximation the following relation:
\begin{equation}
\label{dnu_numax}
\Delta\nu \approx \Delta\nu_\odot \left(\frac{\nu_{\rm{max}}}{\nu_{\rm{max},\odot}}\right)^b \;\; ,
\end{equation}
where $b$ has often values around 0.75. Figure~\ref{nmaxdnu} (top) shows the relation $\Delta\nu$ versus $\nu_{\rm{max}}$ for 1700 stars from the main sequence phase (black diamonds at the right hand side) to the red clump phase (red triangles appearing in a diagonal branch between 20 to 50 $\mu$Hz in the bottom panel) observed by the \emph{Kepler} satellite. Although the relation appears to be constant for all of these stars, several authors \citep{2010A&A...517A..22M,2010ApJ...723.1607H} have suggested that the slope is different between red-giant and main-sequence stars. To enhance such a difference, we have subtracted the luminosity dependence by raising $\nu_{\rm{max}}$ to the power of $b=0.75$. A fit to the residuals below and above $\nu_{\rm{max}}$= 300 $\mu$Hz -- which roughly marks the transition from low-luminosity red giants to sub-giants results in a steeper slope with increasing $\nu_{\rm{max}}$. It is important to note that for $\nu_{\rm{max}}$  close to the solar value, for example, the use of a power-law relation calibrated to red-giant stars would lead to an underestimation in $\Delta\nu$ by $\sim 10\%$.

Very precise surface-gravity values, ($g$), -- with uncertainties much less than 0.05 dex --  can also be inferred directly from the global seismic parameters and from the effective temperature: $g \approx \nu_{\rm max} \sqrt{T_{\rm{eff}}}$ \citep[e.g.][]{2012MNRAS.419L..34M,2012A&A...543A.160T,2013A&A...556A..59H}. This is very important because the seismic $\log g$ values can be used as an independent fixed parameter in the spectroscopic analysis to compute the effective temperatures in an iterative way \citep{2012arXiv1208.6294S}. Indeed, both parameters are correlated and cannot be determined independently from the spectra. Therefore, the combination of a seismic $\log g$ and a spectroscopic $T_{\rm{eff}}$ can reduce the uncertainties in both parameters. CoRoT and \emph{Kepler} stars with asteroseismic surface gravities will be used as calibration stars for the GAIA mission  \citep{2013arXiv1302.7158C}.

\subsection{Stellar Convection and the Background Signal in Oscillation Spectra}

 Solar-like pulsating stars are those in which their acoustic modes are excited by turbulent motions occurring in their convective outer layers \citep[e.g.][]{1977ApJ...212..243G,1992MNRAS.255..603B,1999A&A...351..582H,2005MNRAS.360..859C,2008A&A...478..163B}. The first observations of granulation are quite old \citep{1801RSPT...91..265H}, but it is not until 1933 that their turbulent origin was unveiled \citep{1933AN....247..297S}.

In solar-like stars, exhibiting outer convective envelopes, the granules evolve with time producing small photometric fluctuations.  Therefore, the analysis of stellar light curves in the Fourier domain shows a continuous ``background'' on top of which the acoustic modes are superimposed \citep[see some examples of main sequence stars, subgiants and red giants in][]{2010A&A...522A...1K,2010ApJ...713L.169C,2011A&A...534A...6C,2011A&A...530A..97B,2011ApJ...733...95M,2012ApJ...748L..10M,2013A&A...549A..12M}.

The first successful attempt to model this continuous background was done by \citet{harvey85}. He approximated each scale of the convective motions (granulation, supergranulation...) by an exponentially decaying time function corresponding to the autocovariance of its time evolution with two parameters: the characteristic time of the granules and the total brightness fluctuations associated with them \citep[see][for a complete review of the different formulations commonly used in asteroseismology]{2011ApJ...741..119M}. Moreover, it has  been demonstrated very recently that the properties of the granulation and the global seismic variables are related \citep[e.g.][]{2011ApJ...732...54C,2011ApJ...741..119M}. It is also important to notice that in some stars another feature is visible close to the p-mode envelope that has been related with the presence of faculae in the star \citep[][and references therein]{2013arXiv1302.5563K}. Recent theoretical work by \cite{2013A&A...559A..40S} has also tried to explain the observed relationship between granulation amplitude in the oscillation spectrum and stellar spectral type.

In the context of stellar dynamics, putting observational constraints in the convection inside stars is very important because convection is one of the major ingredients to understand stellar magnetism and dynamo processes. Therefore, precise time domain measurements can offer direct constraints on the granulation characteristics as well as the precise determination of the convective outer envelope. Recently, using \emph{Kepler} data, it was for example possible to calibrate the mixing-length parameter $\alpha_{\rm MLT}$ used in classical 1-D treatment of convection, and to show that the solar value, usually used for generating grids of stellar models \citep{2012A&A...537A.146E} is overestimated \citep{2012ApJ...755L..12B}. Likewise granulation data can put constraints on local \citep{2013A&A...557A..26M} and global \citep{2012ApJ...757..128M} stellar convection simulations, that could in turn help us to characterize stellar oscillations, and convection parameters such as the mixing length \citep{2011ApJ...731...78T}. This important by-product of asteroseismology studies can help to better constrain convective heat transport in 1-D stellar models, and hence to constrain the stellar radius which depends on $\alpha_{\rm MLT}$.

Asteroseismology can also provide important information on the depth of convective envelopes. At the base of the surface convection zone, there
is an abrupt change of stratification (going from adiabatic to subadiabatic) which impacts the sound speed and density profiles (see the pionnering work on the base of the solar convection by \citep{1991ApJ...378..413C}. By using appropriate seismic diagnostic techniques it is then possible to infer the extent of the convective zone of solar-like stars \citep{1990LNP...367..283G,2004A&A...423.1051B,2004AIPC..731..193H,2007MNRAS.375..861H}. This approach has now been successfully applied to \emph{Kepler} data for field stars \citep{2014ApJ...782...18M}.

Complementary to the analysis of asteroseismic data, numerical simulations to model and describe the turbulent state of stellar convective enveloppes or cores have been pursued. Several groups have undergone a systematic study of the global properties of stellar convections and their associated mean flows (differential rotation and meridional circulation). We refer to \S 2 for a more detailed discussion of numerical simulations of rotating and magnetized convective stars.

\subsection{Stellar Rotation and Spin Down}

All stars rotate, and understanding their rotation history is key to our overall understanding of their dynamics.  The angular momentum evolution of stars is an initial value problem, and a wide range of rotation rates are observed in the youngest protostars \citep[see for example][for the Orion Nebula Cluster]{2006ApJ...646..297R}. The relatively slow rotation of such stars \citep{1981ApJ...245..960V} demands efficient angular momentum loss from the infalling material during the hydrodynamic collapse phase, probably through strong magnetic interactions \citep{1994ApJ...429..781S}. Gaseous accretion disks persist around protostars for up to 12 Myr \citep{2013MNRAS.434..806B}, and the exchange of angular momentum between them and their disks can dramatically modify their evolution.  In the limit of strong magnetic interaction, star-disk coupling can regulate stellar rotation, preventing protostars from spinning up as they contract \citep{1991ApJ...370L..39K,1995A&A...294..469K} but see also \citep{2012ApJ...754L..26M}.  The early angular momentum evolution of stars is therefore a window into the star and planet formation process.  Stars also reach the main sequence with a range of rotation rates.  This can cause star-to-star differences in the degree of rotationally induced mixing \citep[see][for a review]{1997ARA&A..35..557P}.  Once on the main sequence, the surface rotation of stars responds both to internal angular momentum transport and to angular momentum loss from magnetized solar-like winds.  There are two strong practical applications that result:

\begin{itemize}
\item 1) We can use the observed angular momentum evolution as a test of stellar physics. Many instabilities and physical processes such as dynamo action, mixing, large scale flows are directly connected to the rotational rate of stars.  Stars differ in their thermal structure, rotation rates, and lifetimes; we therefore have natural laboratories for testing physical models.  This is especially important for our understanding of solar-like winds, as angular momentum loss driven by them is a central driver of angular momentum evolution in low mass stars.
\item 2) With calibrated models we can then use rotation as a stellar population diagnostic.  \cite{1967ApJ...148..217W} explained the slow solar rotation as a consequence of angular momentum loss from a magnetized solar-like wind.  They predicted ${\rm d}J/{\rm d}t \sim \Omega^3$, which implies $\Omega \sim t^{-0.5}$ in the asymptotic limit.  \cite{1972ApJ...171..565S} confirmed this time dependence in rotation, chromospheric activity, and lithium depletion in open cluster stars.  Rotation is therefore a potential age indicator, provided both that it can be measured and that the rotation-mass-age relationship can be calibrated.
\end{itemize}

Both our observational knowledge and the sophistication of our theoretical models have advanced considerably since this pioneering work.  We therefore briefly summarize the observational picture (see Figure 3) before proceeding to the theoretical picture that has emerged. Rotation rates can be inferred spectroscopically from Doppler line broadening.  Such measurements, however, suffer from projection effects ($v \sin i$ is what is measured) and become difficult when the rotational broadening becomes comparable to or smaller than other sources, such as microturbulence or thermal broadening.  Rotation velocity measurements are therefore primarily used for young and massive stars, and through the 1980s and early 1990s most rotation data was spectroscopic. In a pioneering paper, \cite{1987ApJ...318..337S} demonstrated that cool stars in the Pleiades arrived on the main sequence with a wide range of rotation rates, which created three significant problems for simple angular momentum evolution models.  Extremely large torques would be predicted for rapid rotators \citep{1990ApJS...74..501P}, and an unsaturated wind law would not permit the survival of high rotation rates onto the main sequence.  Magnetized solar-like winds are too inefficient to spin down young stars to the observed rates for the slowest rotators, requiring an additional mechanism for extracting angular momentum in the pre-main sequence.  Finally, once on the main sequence, there is a rapid initial drop in rotation rate, followed by a pause where surface rotation changes little; older stars then resume spinning down at the predicted asymptotic rate.  This is evidence for core-envelope decoupling \citep{MacGregorBrenner1991,1997ApJ...480..303K} on timescales of tens to hundreds of Myr \citep[see][for recent discussions]{2010ApJ...716.1269D,2013A&A...556A..36G}. The observed pattern is therefore richer than a simple power-law relationship, requiring both more sophisticated theory and stronger empirical constraints.

Star spots cause periodic brightness modulations as they transit across the visible disk.  Dedicated time domain studies of stars can therefore measure rotation periods.  Rotation period measurements do not suffer from inclination effects (although they require a favorable inclination angle, e.g. not pole-on, to produce a signal.)  There are, however, significant observational selection effects that favor the detection of short period systems, which also tend to be high amplitude.  There has been an enormous increase in the quantity and quality of stellar rotation period data.  In particular, the advent of large surveys designed to search for extrasolar planets has yielded very large data sets of stellar rotation periods as a natural byproduct for both field stars \citep{2011AJ....141..166H} and open clusters stars.  Examples of the latter include the extensive MONITOR program \citep{2007MNRAS.375...29A} \citep[see][for the most recent results]{2013A&A...560A..13M}, the SuperWASP data for the Hyades and Praesepe \citep{2011MNRAS.413.2218D} and the large \citep{2009ApJ...691..342H,2010MNRAS.408..475H} M37 and Pleiades data sets.  The \emph{Kepler} and CoRoT missions reach lower amplitudes and can detect stars with longer rotation periods, pushing stellar rotation measurements into the solar regime \citep{2011AJ....141...20B,2014arXiv1403.7155G,2013ASSP...31..215L,2013MNRAS.436.1883W,2013A&A...557L..10N,2013arXiv1308.1508R,2014ApJS..211...24M}.
With the aid of these samples, a standard empirical picture of angular momentum evolution has emerged, with the following basic ingredients.

\begin{figure}[!htb]
\begin{center}
\includegraphics[width=0.9\textwidth]{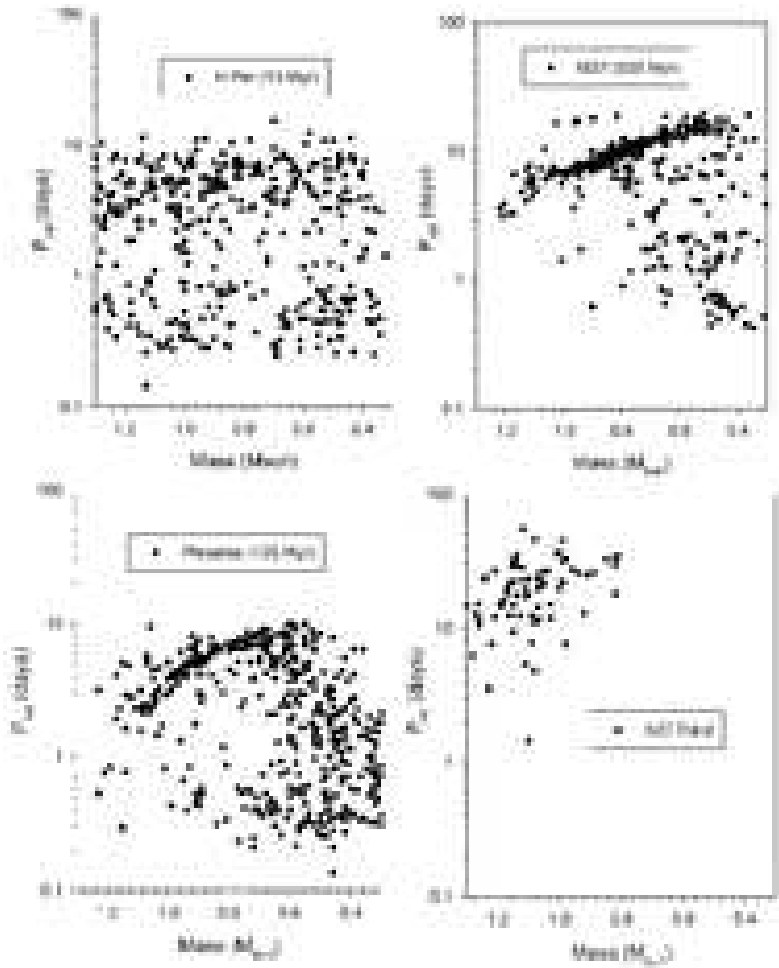}
\end{center}
\caption{\label{obstrendrot} The observed distribution of rotation rates in stars of different ages.  Data for the young protostars in H Per (upper left) is taken from \cite{2013A&A...560A..13M}.  The data for the 125 Myr old Pleiades system (lower left) is taken from the reanalysis by Coker \& Pinsonneault (2015, in prep).  M37 data is originally from \cite{2009ApJ...691..342H} as interpolated onto mass by \cite{2013ApJ...776...67V}.  Data for old field dwarfs is taken from \cite{2014arXiv1403.7155G}.}
\end{figure}

Stars are born with a range of rotation rates, and they would be expected to spin up as they contract and approach the main sequence.  Through protostar-disk interactions some stars lose additional angular momentum in the first $\sim$10 $Myr$, widening the observed distribution.  The slow rotator population cannot be generated from the young star forming regions, as noted above, but is already in place in the 13 Myr old system H Per \citep{2013A&A...560A..13M}. This confirms the hypothesis that protostar-disk interactions are important for stellar angular momentum evolution.  The distribution of rotation rates is initially broad and narrows with age, which is a natural consequence of a torque that increases more than linearly with rotation rate.  This convergence process is most rapid for solar mass stars and the timescale for spin down increases with decreased mass.  This is usually interpreted as evidence for mass-dependent saturation in the angular momentum loss rate, typically parameterized with a Rossby scaling (rotation period relative to the convective overturn timescale.)  Core-envelope decoupling is required to explain young slow rotators, with a time scale that increases with decreased mass \citep{2010ApJ...716.1269D,2011MNRAS.413.2218D}.  There is no break or discontinuity in the observed spin down behavior at the fully convective boundary, which is surprising given dynamo model predictions.  New data sets in older solar analogs in \emph{Kepler} stars will provide tests of the mass and composition dependence of rotation in old dwarfs, which is currently lacking. Field star studies will need to consider the presence of background populations when modeling angular momentum evolution.  For example, angular momentum lost from tidally synchronized binaries is extracted from the orbital rather than the spin angular momentum reservoir, which results in very different angular momentum evolution properties \citep[see][for a discussion in the context of blue stragglers]{2006ApJ...646.1160A}.

These large data sets have spurred renewed interest in the usage of stellar rotation as an age indicator, or gyrochronology \cite{2003ApJ...586..464B}.
Distinct theoretical scalings and fitting function have been proposed \citep[e.g.][for a discussion]{2008ApJ...687.1264M,2010ApJ...721..675B,2014ApJ...780..159E}. There are significant zero-point shifts between these formulations, which will require resolution when compared with empirical data.  There are strict lower bounds to the uncertainties in gyrochronology relations even for an ideal theoretical framework.  \cite{2014ApJ...780..159E} identified two fundamental limiting effects for the precision of gyrochronology: The intrinsic range of rotation rates at fixed age and the range of surface rotation rates in a given star from latitudinal differential rotation.  The former is more important for young and low mass stars, while the latter dominates for older higher mass stars.  The net effect is an age uncertainty at the 2 Gyr level for old stars.  \cite{2013ApJ...776...67V} extended this to subgiants, and argue for a different gyrochronology relationship for them than for dwarfs.
Recently the gyrochronology concept  has been adapted  to consider magnetic field instead of rotation as a proxy for age: e.g. {\it magnetochronology} \citep{2014arXiv1404.2733V}.
It is indeed clear (see next section) that stellar magnetic properties vary over the course of stellar evolution, older stars being less active their younger counterparts \citep{2005ssac.conf..307G,2012ApJ...755...97G}.

The theoretical interpretation, numerical models and multi-D simulations allowing us to study and to understand the stellar age-rotation-activity relationship will be discussed in \S 2.\\

\subsection{Stellar Activity \& Magnetism}

The great advantage of comparing the Sun to other stars is to disentangle what is specific to our star from what
is generic to solar-like stars. Systematic observations of rotation and magnetic properties of solar-like stars have revealed interesting trends.
For instance surface differential rotation is found to increase with $T_{\rm eff}$; F-star possess a larger latitudinal contrast than K-stars do  \citep{1996ApJ...466..384D,2005MNRAS.357L...1B,2007AN....328.1030C}. Similarly a rotation-activity relationship has been known for quite some time now \citep{2003A&A...397..147P}.
A relation between cycle and rotation periods has also been advocated \citep{noyes84a,2007ApJ...657..486B}.

Meaningful information regarding magnetic cycles underlying stellar activity, or alternatively to rule out the existence of magnetic cycles, require long-term stellar activity observations and their analysis. Direct observations of starspots is a relatively new capability and studies of their behavior over timescales long enough to be relevant for stellar dynamos, do not yet exist. Nonetheless, indirect evidence of magnetic activity cycles (or the lack thereof!) can be obtained by different techniques. Our traditional tools for measuring stellar cycles have been synoptic stellar chromospheric activity observations, complemented with more limited stellar coronal X-ray variability data.  Asterseismology may prove a useful additional set of data.

The most important contribution in the context of long-term stellar activity observations has been the seminal Mount Wilson Observatory (MWO) Calcium (Ca) H+K Project (\citealt{wilson78, noyes84a, baliunas85, baliunas95}).
The emission in these line cores, expressed as a Ca H+K activity index,  is understood to be a measure of the magnetically mediated, non-thermal heating of the chromosphere \citep[provided that the basal thermal flux is properly accounted for, e.g.][]{1989ApJ...341.1035S,1999ApJ...522.1053C}. Thus, the long-term variation of the Ca H+K index can be taken as a proxy for variability in stellar magnetic fields (see the review by \citealt{hall08}). The Lowell Observatory's Solar-Stellar Spectrograph program provides complementary observations of Sun-like stars (\citealt{hall07,hall08,2009AJ....138..312H}) adding to the knowledge base of the MWO program. An independent measure of stellar magnetic output is the coronal X-ray flux -- which originates in the magnetic heating of stellar coronae and which is even more tightly correlated with magnetic flux (\citealt{pevtsov03}). Although stellar X-ray cycle observations are rare, the amplitude of stellar X-ray flux have been observed for multiple stars (\citealt{hempelmann96, micela03, gudel04}) and this usefully constrains the strength of the dynamo generated magnetic fields. More recent surveys analyzing the activity of solar-like stars are to be found in \cite{wright04} and \cite{giampapa06}. We provide here a brief summary of the important trends and relationships relevant for stellar activity cycles that have been gleaned from these long-term observations.

\subsubsection{General Activity Trends with Rotation}
 First we focus on the scaling behavior of magnetic activity with the stellar rotation rate. Stars that are faster rotators have higher magnetic activity amplitudes than slower rotators.  They are usually variable, but do not display regular cycles. Stars with intermediate rotation rates have intermediate magnetic activity output, which sometimes display periodic behavior. Stars which rotate relatively slowly, such as the Sun, are more likely to have regular, periodic cycles, but their activity amplitudes are also relatively smaller. The latter category also sometimes display``flat'' (i.e., non-varying) activity with low amplitude, akin to global magnetic minima such as the solar Maunder minimum. We note here that fast rotators also sometimes display a ``flat'' (non-varying) activity profile albeit with high activity amplitudes-- which is not to be confused with global minimum like states.

\begin{figure}[!htb]
\begin{center}
\includegraphics[width=0.7\textwidth]{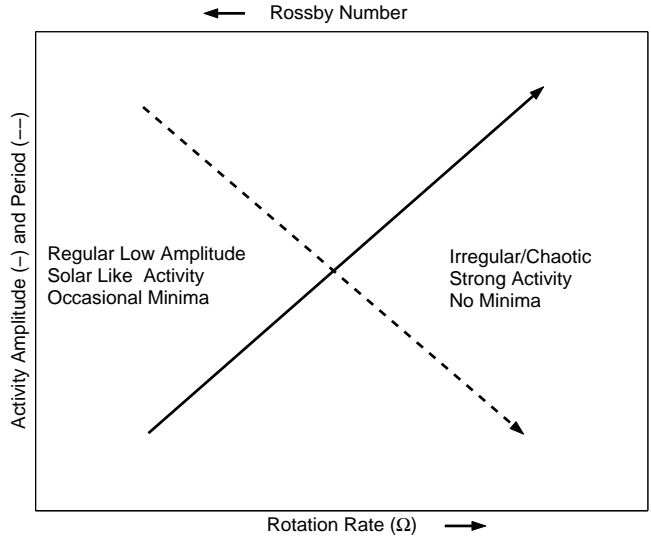}
\end{center}
\caption{\label{obstrendsmag} Stellar activity trends as determined from observations of
Ca H+K and X-ray emission from F-G-K stars. The arrows in the axis
label show the direction of increase of the rotation rate and Rossby
number. In general, magnetic cycle amplitude increases while cycle
period decreases with increasing rotation rate (i.e., decreasing
rotation period). The critical reader is referred to the original
observations cited in the text for the exact scaling laws and other
subtleties in the data. From \cite{nandy04}.}
\end{figure}

In general the activity amplitudes increase with rotation rates, but for very fast rotators, the measured amplitudes tend to reach a saturation \citep{2003A&A...397..147P,2011ApJ...743...48W}, which in X-rays is of order $\log(L_x/L_{bol}) \sim -3.2$. For G type stars this saturation is found for rotation rate above 35 ${\bf \rm km s^{-1}}$,  for K type stars at about 10 ${\bf \rm km s^{-1}}$ and for M dwarfs around 3-4 ${\bf \rm km s^{-1}}$. How stellar magnetic flux scales with rotation rate is important to understand, since it is telling us how the magnetic field generated by dynamo action inside the stars emerges and imprints the stellar surface.  Understanding the apparent saturation phenomenon is also important: for example, it could arise if only the filling factor $f$ reaches a maximum \citep{2012LRSP....9....1R}.  Multi-periodic cycles are also sometimes observed, especially in faster rotators \citep{2009A&A...501..703O}.  For those stars with cyclic magnetic activity, the systematic analysis of stellar data revealed that for solar type stars there is a good correlation between the cycle and rotation periods of the stars and that correlation is even stronger when using the Rossby number ($Ro = P_{rot}/\tau$) that takes into account the convection turnover time $\tau$ at the base of the stellar convective envelope \citep{noyes84a,noyes84b,1993ApJ...414L..33S,1996ApJ...460..848B}. As the star rotates faster, its cycle period is found to be shorter. This trend is summarized in Figure \ref{obstrendsmag}. More precisely, \cite{noyes84b} found that $P_{cyc} \propto P_{rot}^n$, with $n = 1.25 \pm 0.5$. \citet{1999ApJ...524..295S,2002ASPC..277..311S} have used an extended stellar sample to argue that there are actually two branches when plotting the cycle period vs the rotation period of the stars: the primary (starspot) cycle and the Gleissberg, or grand minima type, modulation of stellar activity. For the active branch they found an exponent $n \sim 0.8$ and for the inactive one they inferred $n\sim 1.15$.

\subsubsection{Maunder Minima Stars:}
Many of the stars (roughly 15\%, see \citealt{giampapa06}), studied in the Mount Wilson surveys, displayed activity that were flat (i.e., non-cyclic) and at or below current solar minimum levels.
Similar studies with \emph{Kepler} data recently found that among the sampled solar-like seismic stars about 2\% show a flat activity level.
One possible interpretation of these results is that these are ``Maunder minima'' stars where the global magnetic activity has temporarily ceased. Complementary studies of coronal emission also point to the existence of such low-activity stars (\citealt{poppenhager09}). \cite{baliunas95} also provides evidence indicative of a star entering into a Maunder minimum like phase -- with its activity variation flattening out accompanied by a reduction in amplitude.

\subsubsection{Starspots, Polar Caps and Stellar Butterfly Diagrams:}
Recent advances in stellar photometry and doppler imaging techniques now allow us to observe starspots and their dynamics. Such observations (reviewed in \citealt{berdyugina05}) show that there is oftentimes a preferred longitude for starspots. Faster rotators tend to have large polar spots which contributes  to a strong polar cap, while in general, in very active stars, the spot distribution tends to be more uniform across all latitudes. In some stars, the spots evidently appear at mid-low latitudes and subsequently migrate to higher latitudes, reminiscent of the sunspot migration (although towards low latitude) that generates the solar butterfly diagram. These observations do not span a long period of time and hence detailed studies of stellar butterfly diagrams is still in its infancy. Nevertheless, observations of spot emergence latitudes and migration are expected to provide more stringent constraints in the future for stellar dynamo and flux emergence models.

\subsubsection{Asteroseismology and Activity}

Ensemble asteroseismology can now add to our developing understanding of stellar activity obtained by other methods, such as spectropolarimetry and CaII H \& K studies.

The first contribution of  helio and astro-seismology comes through their ability to provide stellar properties and, in particular, stellar ages for field stars that are otherwise difficult to analyze. Our lack of knowledge of stellar ages has been an important limitation in our understanding of the Mt. Wilson survey data, for example.  Many of the solar-like stars showing a flat S-index in the original sample analyzed by  \citet{1990Natur.348..520B} --and assumed to be in a similar state to the Maunder minimum in the Sun--  were found to be more evolved stars (which would be expected to less active) when accurate parallaxes from the Hipparcos mission were taken into account \citep{2004AJ....128.1273W}. Therefore, accurate stellar properties are essential for proper interpretation of their magnetic indexes \citep[see an extended discussion in the review by][]{2012IAUS..286...15J}. But seismology can also be employed to directly study stellar activity. Magnetic fields modify the conditions inside the stars and the properties of the modes are changed accordingly. Seismology can therefore provide complementary information to the one obtained by the above mentioned techniques. Moreover, it can for example provide evidences of the existence of subsurface on-going magnetic cycles even during Maunder minimum-like states \citep{1998SoPh..181..237B} in which no external manifestation of the internal magnetism can be observed. However, the amplitude of the modes decreases for active stars as observed in the Sun \citep[e.g.][]{2003ApJ...595..446J}, HD~49933 observed by CoRoT \citep{2010Sci...329.1032G},  and solar-like stars observed by \emph{Kepler} \citep{2011ApJ...732L...5C} difficulting their observation.

The first diagnostic or ``proxy for magnetic activity'' used in helioseismology was the frequency shift of the acoustic modes, i.e., the displacement of the modes towards higher frequencies as the cycle evolves. These frequency shifts --which follow an inverse mode-mass scaling \citep{1990Natur.345..779L}-- are explained to arise from changes in the outer layers of the Sun along the 11-year magnetic cycle (Schwabe cycle) \citep{1991ApJ...370..752G,1996MNRAS.278..437B,2004ApJ...600..464D,2005ApJ...625..548D}. Therefore, the high-frequency p modes -- having higher upper turning points than the low-frequency p modes -- are then more sensitive to the perturbations induced by the magnetic field in the outer part of the Sun. Thus, these high-frequency p modes are observed to have large frequency shifts with the Schwabe cycle, while the low-frequency p modes present little or no variation with it \citep[e.g.,][]{1994SoPh..150..389R,1998MNRAS.300.1077C,2002A&A...394..285G,2008ASPC..383..305H,2009ApJ...695.1567J}. A similar behavior has also been found in the F star HD~49933  \citep{2011A&A...530A.127S}.

It is now commonly accepted that, apart from the rotational splitting \citep{2008SoPh..251..119G,2012MNRAS.422.3564B}, all the rest  of the solar low-degree p-modes properties change with time \citep[e.g.][]{2000MNRAS.313...32C,2003ApJ...595..446J,JimCha2007,2011JPhCS.271a2030S,2013JPhCS.440a2020G} in a time scale of 11 years following the corresponding surface Schwabe cycle. Another shorter timescale has been recently confirmed using helioseismology. Indeed, the quasi-periodic modulations of 1 to 2 years observed during the last twenty years in the Sun \citep[e.g.][]{1995SoPh..161....1B,2003SoPh..212..201M} has then been measured \citep{2010ApJ...718L..19F} in the frequency shifts of low-degree modes measured by the Global Oscillation at Low Frequencies \citep[GOLF][]{GabGre1995}) instrument on board the Solar and Heliospheric Observatory \citep[SoHO][]{DomFle1995}, as well as with the Birmingham Solar Oscillation Network \citep[BiSON][]{1996SoPh..168....1C}. Although the origin of this modulation is not  well understood yet,  the latest results priviledge the hypothesis that the origin of the perturbation is located in the subsurface layers \citep{2013ApJ...765..100S}.

The continuous photometric monitoring of hundreds of thousands of stars by CoRoT and \emph{Kepler} provide long baseline stellar variability samples of unprecedented scale and precision.  Their duration is also substantial, comprising a period of more than 4 years in the case of \emph{Kepler}; this is long enough to begin to see the effect of activity cycles. The first ensemble analysis \emph{Kepler} light curve varibility was done by \citet{2010ApJ...713L.155B} and \citet{2011AJ....141...20B}. However, because the \emph{Kepler} pipeline was not designed to measure long-period stellar variability they used raw data from only a 90 day quarter, which introduced important limitations in their analysis. \citet{2011AJ....141...20B} found that the amplitude of variability was in general larger for stars that were clearly periodic than for those that were not. As expected, the largest group of non-periodic variables were giant stars. To improve on the \emph{Kepler} pipeline, which was designed for transit detection and not stellar variability, several groups developed their own correction software  \citep[e.g.][]{2011MNRAS.414L...6G,2014ApJS..211...24M,2014arXiv1403.7155G}. With these new analysis tools it was possible to detect longer-than-a-month rotation periods as well as to study the magnetic variability of the stars.

\begin{figure}[!htb]
\begin{center}
\includegraphics[width=0.85\textwidth]{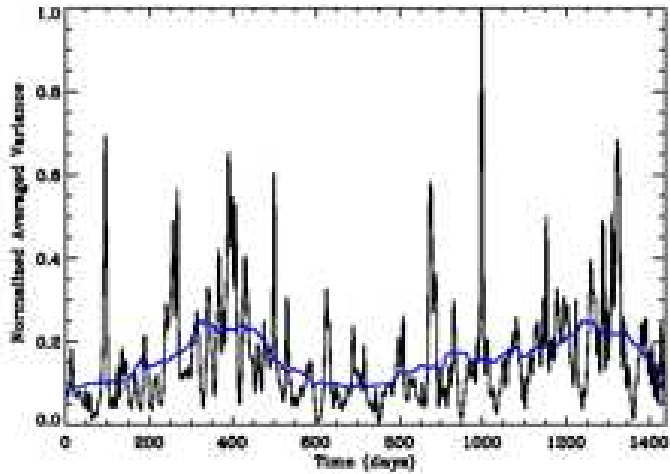}
\end{center}
\caption{\label{FigMat13}Evolution of the normalized averaged variance of the \emph{Kepler} F star KIC~12009504 for which p-mode oscillations have been measured
(in black) and smoothed by a boxcar function (160 days wide) in blue. A regular cycle of around 800 days is observed \citep[see][for further details]{2014A&A...562A.124M}.}
\end{figure}

Based on broadband photometric variability, a new proxy for stellar magnetic activity has been developed \citep{2010Sci...329.1032G,2011ApJ...732L...5C,2014JSWSC...4A..15M}. The time evolution of the variance of the signal appears to be a good indicator of the magnetic activity cycle. It has been successfully tested on the Sun using VIRGO/SPM data recorded during the last 16.5 years \citep[see for more details][]{2013JPhCS.440a2020G}. Using the same methodology, \citet{2014A&A...562A.124M} studied a subset of 22 \emph{Kepler} F stars with rotation periods of less than 12 days for  which p-mode oscillations were reported. Because of the relation between the rotation period of the star and the period of the cycle \citep[e.g.][]{2007ApJ...657..486B,2010A&A...509A..32J}, they were expecting to see in some stars cycle-like variations such as the one observed by \citet{2010ApJ...723L.213M} in the F8V star $i$~Horlogii. That star exhibited a magnetic cycle period of 1.6 years for a rotation period of 8.5 days. \citet{2014A&A...562A.124M}  found that only a small fraction of stars showed a regular cycle-like behavior (see Fig.\ref{FigMat13}) and there were no clear trends between activity and rotation in the F-star sample. A more systematic study still needs to be done to confirm or refute the existence of a general relationship between stellar cycle period and stellar rotation period for these hot stars. For solar analogs recent analysis of \emph{Kepler} data have confirmed an rotation-age relationship \citep{2014ApJ...790L..23D}.

\section{Global Models of Stellar Structure, Rotation, and Magnetism}

In order to understand, characterize and describe the physical processes at the origin
of stellar convection, rotation and magnetism, with the Sun being an archetype of solar-like stars, we must
develop a self-consistent, ideally nonlinear, description of their dynamics. Depending on the time scale of interest and given the current
computing resources such model can be either 1-D, 2-D or 3-D. In the following section we briefly summarize the recent
progress made in modeling the multi-scale dynamics of stars and their success at explaining the observed properties discussed in \S 1.

\subsection{1-D stellar evolution and core-envelope coupling}

The angular momentum evolution of a star - its rotational history from its birth until its death Ð has profound consequences for its structure, evolution, and dynamics.  Rotationally induced mixing for low mass stars can induce mild mixing that alters surface abundances (see \citealt{1997ARA&A..35..557P} for a review), while for massive stars, which rotate much more rapidly, the impact of mixing is far more dramatic \citep[see][for a recent review]{2012RvMP...84...25M}. Stellar rotation is also intimately connected with the origin and generation of stellar magnetic fields, and the two together shape wind outflows from stars.  Hence it is fundamental to be able to characterize it.  The overall picture of angular momentum evolution was discussed in section 1.3.  Here we focus on modeling.

Structural evolution in stars will naturally generate internal angular velocity gradients, and in some evolutionary stages these shears can be large.  Mass loss from the surface will also carry angular momentum away from the star, providing another powerful driver for radial differential rotation.  In convective regions the interaction of rotation, magnetism, and convection will enforce an angular momentum distribution on a convective overturn timescale, much shorter than the natural nuclear evolution timescale.  The response of radiatively stable regions is more complex, both because of our limited knowledge of the physics and because of limited empirical constraints on the relevant time scales.  Internal redistribution by hydrodynamic and magnetohydrodynamic physical process can both be important, and different classes of models testing different prescriptions have been employed.

Solar-like stars on the main sequence evolves on long time scales; the solar main sequence lifetime is of the order 10 Gyr. Computing such
long temporal (secular) evolution with multi-D codes is impractical. One must rely on 1-D stellar evolution models to describe the structure, chemical and rotational evolution of stars. We refer to the recent review by \citep{2014SSRv..tmp....3B} for a detailed discussion of 1-D stellar evolution codes, their success and difficulty at reproducing the solar internal structure, the influence of chemical composition, and how such helioseismically calibrated 1-D solar models have been extended to compute up-to-date 1-D stellar models. Here we focus on internal angular momentum transport and its consequences for the rotation history of stars. To this end, let's apply the concept of angular momentum exchange and conservation to a star like our Sun.

\begin{figure*}[!htb]
\begin{center}
\includegraphics[width=0.4\textwidth]{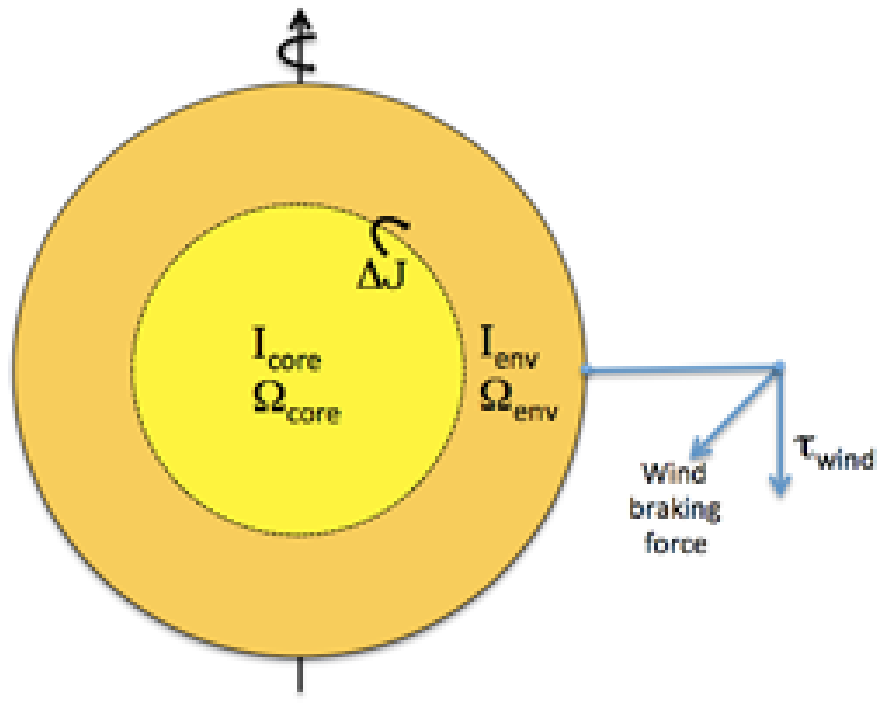}
\includegraphics[width=0.59\textwidth]{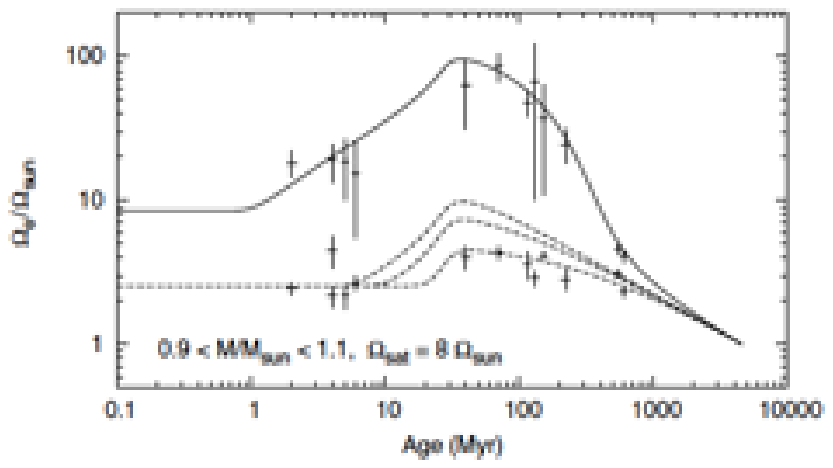}
\end{center}
\caption{\label{amon_hist} Left: The two layers model of a stars, showing the exchange of angular momentum $\Delta J$ between a core and an envelope rotating at different rates and
the braking action of a stellar wind. Right: Evolutionnary track for two layer models showing how the coupling time scale can explain slow and fast rotators (from \cite{2010ApJ...716.1269D})}
\end{figure*}

A useful framework that has been extensively employed in the literature is the two zone model of  \cite{MacGregorBrenner1991}. In this model, the solar-like star is treated as a sphere divided into two zones: an inner stably stratified core and a turbulent convective envelope both independently rotating around an axis aligned with the $\hat{\bf e}_z$ direction, where $\hat{\bf e}_z$ is the unit vector along the z axis of a 3-D $(x,y,z)$ cartesian system (see Fig \ref{amon_hist} left panel). Each possess their own moment of inertia $I$ and angular velocity $\Omega$; both structural evolution and magnetized winds can generate differential rotation. Their angular momentum is respectively $J_{core}=I_{core}\Omega_{\rm core}$ and  $J_{\rm env}=I_{\rm env}\Omega_{\rm env}$. Let's now find the exact quantity of angular momentum exchange $\Delta J$ between the two zones require to have them rotate uniformly at the rotation rate $\Omega$. The initial state is: $J_{\rm init}^c= J_{\rm core}$
and $J_{\rm init}^e= J_{\rm env}$. Applying total angular momentum conservation e.g. $J_{\rm init}^c + J_{\rm init}^e = J_{\rm final}^c + J_{\rm final}^e$,
implies that the final state is: $J_{\rm final}^c= J_{\rm core} - \Delta J = I_{\rm core} \Omega$
and  $J_{\rm final}^e= J_{\rm env} + \Delta J = I_{\rm env} \Omega$. A simple substitution yields:
 \begin{equation}
\Delta J = \frac{I_{\rm env}J_{\rm core}-I_{\rm core}J_{\rm env}}{I_{\rm core}+I_{\rm env}}.
\end{equation}
If an external torque, such as the one coming from a stellar wind $\tau_{wind}$ is applied to the star,
the angular momentum evolution of the convective envelope and radiative interior can be expressed as \citep{MacGregorBrenner1991}:
\begin{eqnarray}
\frac{dJ_{\rm core}}{dt} &=& - \frac{\Delta J}{t_{\rm c}}  \\
\frac{dJ_{Örm env}}{dt} &=& \frac{\Delta J}{t_{\rm c}} - \frac{J_{\rm env}}{t_{\rm w}} \, ,
\end{eqnarray}
with $t_{\rm c}$ the coupling time scale between the two zones due to the combined action of turbulence, waves, magnetic fields and
viscous stresses and $t_{\rm w}$ the wind braking timescale. These formula further assumes that the convective envelope transmits instantaneously the
applied surface torque to the base of the convective envelope.

Such models have been used to explore the relevant coupling time scale between
the convective envelope and the radiative interior in solar-like stars over the course of their evolution \citep{MacGregorBrenner1991,1995A&A...294..469K,1997ApJ...480..303K,Allain1998} and \cite[for recent developments see][]{2010ApJ...716.1269D,2013EAS....62..143B,2013A&A...556A..36G,2013ApJ...778..166O,2014ApJ...787..131Z}.  It is found that a time scale of tens to hundreds of Myr can explain the core-envelope coupling in young open cluster stars (see Figure \ref{amon_hist}).

One can identify key angular momentum transport processes which are expected to act inside a stellar convection zone or radiative interior: Reynolds and Maxwell stresses, meridional circulation, viscous processes (usually negligible), large scale magnetic torque, gravity waves \citep[see][and references therein]{1989ApJ...338..528G,1992A&A...265..115Z,1997ARA&A..35..557P,2004A&A...425..229M,2005A&A...440..653M,2010ApJ...716.1269D,2011LNP...832..275M,2011ApJ...742...79B,2012ApJ...746...43R,2013LNP...865...23M,2013A&A...556A..36G,2013ApJ...776...67V}. The amplitude and role of each process varies as a function of time. For example, it is known that young stars are more active than old ones, so internal Maxwell stresses are expected to play a less significant role later in the star's evolution on the main sequence and likewise for the stellar wind torque. Of course when the star undergoes a large structural change due to its secular evolution (as for instance when it goes up the red giant branch) the relative importance of different processes must be re-evaluated.

\cite{1988ApJ...333..236K} proposed a general analytical formulation of the applied torque that has been widely used. He adopted the \citep{1968MNRAS.138..359M} framework and adopted parameters tuned to reproduce Skumanich's law \citep{1972ApJ...171..565S} for the time dependence of surface rotation.
Many other studies have since improved upon Kawaler's model by taking into account better observational information or theoretical improvements in order to assess how the magnetic torque applied by a stellar wind depends on stellar parameters and dynamics (rotation, dynamo); see for example the Rossby-scaled saturation threshold adopted by \cite{1997ApJ...480..303K}, which has been extensively employed subsequently.  There has been much recent activity in this area, with new frameworks being proposed by (\citealt{2011ApJ...741...54C,2012ApJ...754L..26M,2012ApJ...746...43R,2013ApJ...776...67V,2014ApJ...789..101B,matt14}). \\

Self-consistent solar models including angular momentum transport from hydrodynamic mechanisms were first generated in the 1980s \citep{1981ApJ...243..625E,1989ApJ...338..424P}.
Meridional circulation, hydrodynamic instabilities and gravity waves are the primary mixing and angular momentum transport agents, so these models play an important role in rotational mixing studies (\citealt{2013EAS....63..269M,brun14}).  Such models are sometimes coupled with approximations of magnetic effects, such as the Taylor-Spruit mechanism \citep{2002A&A...381..923S}. This approach has been used to model massive rotating stars over secular time scales \citep{2003A&A...399..603P,2013A&A...549A..74M,2013AN....334..168M} and to generate grids of stellar models used for isochrones and chemical evolution studies \citep{2012A&A...537A.146E,2013A&A...558A.103G,2012A&A...543A.108L}. Some attempts have also been made to model in 2-D the secular evolution of internal angular momentum by considering the role of large scale magnetic field in the radiative interior \citep[see][and references therein]{1993ApJ...417..762C,1996ApJ...466.1078R,2010MNRAS.404..641S}. However such models do not solve for dynamo action as we will now discuss in the next section.

\subsection{2-D mean field stellar dynamo models}

On time scales of centuries to millenia it is possible to model in 2-D the magnetohydrodynamics of stars using the mean field dynamo framework by parameterizing or neglecting some of the detailed description of convective motions. This allows one to explore the underlying physics and a wide range of stellar parameter space which we describe below.

\subsubsection{The Theoretical Framework}

Stellar magnetic cycles involve the generation and conversion of two components of the Sun's magnetic field -- the toroidal component and the poloidal component. Under the assumption of axisymmetry, the stellar magnetic field and plasma flow can be expressed as
\begin{eqnarray}
{\bf B}& =& B_{\phi} {\bf e}_{\phi} + \nabla \times (A\, {\bf e}_{\phi}) \\
{\bf v} &=& r\sin(\theta)\Omega\,{\bf \hat{e}}_{\phi} + {\bf v}_p
\end{eqnarray}
In this axisymmetric decomposition, the first term on the R.H.S.\  of Eqn.~9 is the toroidal component and the second term the poloidal component of the magnetic field (of which $A$ is the vector potential). The first term on the R.H.S.\ of Eqn.~10 signifies the $\phi$-component of the velocity (i.e., angular velocity  $\Omega$), while ${\bf v}_p$ denotes the meridional ($r$-$\theta$) component of the velocity.

\begin{figure}[!htb]
\begin{center}
\includegraphics[width=0.52\textwidth]{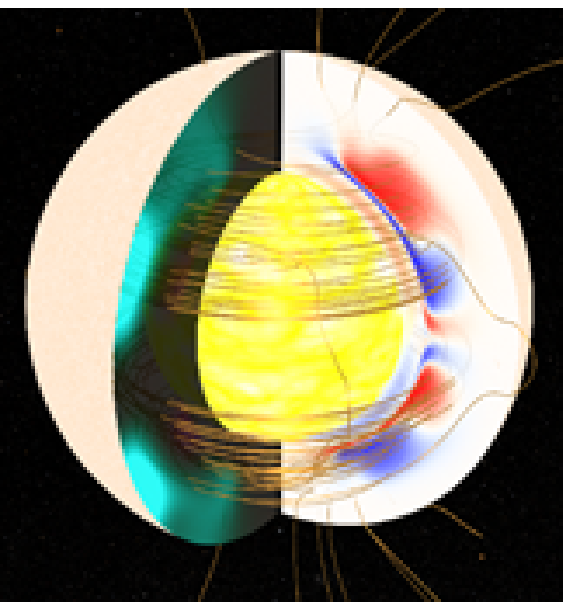}
\includegraphics[width=0.4\textwidth]{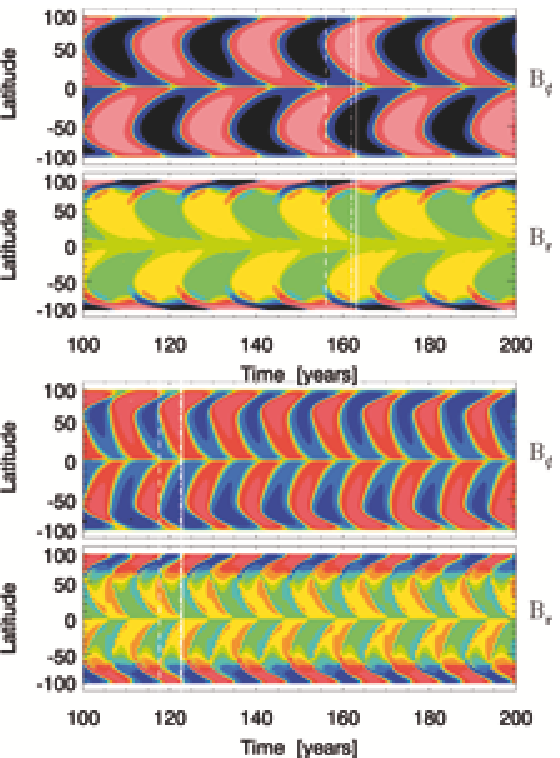}
\end{center}
\caption{\label{blmodel} Left: a 2.5D kinematic dynamo simulation depicting a snapshot
of the toroidal (right meridional plane) and poloidal field
(left meridional place) configurations at a phase close to
solar minimum. The simulation shows that the toroidal component of
the field is generated and stored deep in the convection zone, from
where it buoyantly erupts to form sunspots. The poloidal field is
created in the near-surface layers in this particular dynamo model
of the Babcock-Leighton type \citep{nandy11}. 3-D rendering of field lines (magenta) show
the expected wound-up field structure in the solar interior. Right: Butterfly diagram for 2 models
representative of pumping dominated dynamo, rotating at respectively $0.7 \Omega_\odot$ and $3.0\Omega_\odot$.
We note the equatorward branch, the correct phase relationship between the toroidal and poloidal component of field (shown as vertical lines) and
the shorter rotation period for the case rotating at three time the solar rate \citep{docao12}.}
\end{figure}

Our understanding of solar and stellar cycles is based primarily on 2-D kinematic $\alpha\Omega$ dynamo models and full MHD simulations (cf. \S2.2, \cite{2013SSRv..tmp..100B} (also published as chapter 5 of the ISSI book \citep{sismoissi}) and \citealt{1994lspd.conf...59W,charbonneau10}), most of which have been constructed in the solar context. Here we briefly review this understanding. The toroidal component of the magnetic field is produced by stretching of an initial poloidal field by the differential rotation inside stars (the dynamo $\Omega$-effect) and is thought to be stored in a layer stable to convection near the base of the convection zone. Strong toroidal flux tubes rise up due to magnetic buoyancy and erupt as magnetic (star) spots. The toroidal flux must be recycled to poloidal flux for the dynamo cycle to perpetuate and this conversion mechanism (traditionally termed as the dynamo $\alpha$-effect) could be achieved by various means. The two most popular and widely studied mechanisms are the mean-field $\alpha$-effect driven by helical turbulence (\citealt{parker55}) and the decay and dispersal of tilted bipolar active regions  -- i.e., bipolar sunspot or starspot pairs (\citealt{babcock61, leighton69}), hereby the BL mechanism. It is thought that strong toroidal flux tubes, believed to be necessary to produce sunspots, would quench the mean-field $\alpha$-effect, if there is no segregation of the poloidal and toroidal generation process and thus the BL mechanism must be the dominant mechanism for poloidal field generation in the Sun; however, this debate is far from over and we will return to this later.
We display in Figure \ref{blmodel} a typical representation of a BL like dynamo model.

In the kinematic approach, the plasma velocity is prescribed and the magnetic field evolution equation is solved under the assumption of no direct magnetic feedback on the flows. The magnetic induction equation is given by
\begin{equation}
\frac{\partial \textbf{B}}{\partial t} = \nabla\times\left( \textbf{v}\times\textbf{B} - \eta\nabla\times\textbf{B}\right).
\end{equation}
On substituting Eqn.~9 and Eqn.~10 in the induction equation and separating out the toroidal and poloidal magnetic field components we get the evolution equations for the poloidal and toroidal components, respectively:
\begin{eqnarray}
  \frac{\partial A}{\partial t} + \frac{1}{s}\left[ \textbf{v}_p \cdot \nabla (sA) \right] &=& \eta\left( \nabla^2 - \frac{1}{s^2}  \right)A + S_{\alpha},\\
  & & \nonumber\\
  \frac{\partial B_{\phi}}{\partial t}  + s\left[ \textbf{v}_p \cdot \nabla\left(\frac{B_{\phi}}{s} \right) \right] + (\nabla \cdot \textbf{v}_p)B_{\phi}&=& \eta\left( \nabla^2 - \frac{1}{s^2}  \right)B_{\phi} + s\left(\left[ \nabla \times (A\bf \hat{e}_\phi) \right]\cdot \nabla \Omega\right) \nonumber \\
  & & + \frac{1}{s}\frac{\partial (sB_{\phi}`)}{\partial r}\frac{\partial \eta}{\partial r}.
\end{eqnarray}
Here $s = r\sin(\theta)$ and $\eta$ is the turbulent diffusion coefficient. The source term for the poloidal field $S_{\alpha}$ could denote either the mean-field $\alpha$-effect (in which case $S_{\alpha} = \alpha B_{\phi}$) or the BL source due to the buoyant eruption and flux dispersal of tilted active regions. The flow in the meridional plane ${\bf v}_p$ represents meridional circulation (${\bf v}_p = {\bf v}_m$, where ${\bf v}_m$ denotes the meridional circulation), and in the presence of turbulent pumping (say ${\bf v}_{\gamma}$) it represents the combined effect of both (i.e., ${\bf v}_p = {\bf v}_m + {\bf v}_{\gamma}$). Appropriate prescriptions for the flows, turbulent diffusion and nature and profile of the poloidal field source along with the boundary conditions, completely describes the kinematic dynamo system defined in Eqn.~12 and Eqn.~13.

In BL flux transport dynamo models, the cycle period $P_{cyc}$  is found to depend strongly on the meridional circulation amplitude $v_0$ and its profile and less on the rotation rate $\Omega_0$ or the amplitude of the surface source term $s_0$ \citep{1999ApJ...518..508D,2007A&A...474..239J}:
\begin{equation}
  P_{cyc} \propto \Omega_0^{0.05} S_0^{0.07} v_0^{-0.83}
\end{equation}

As will be seen in the next section, the meridional circulation is found to decrease with the rotation rate as $v_0 \propto \Omega_0^{-0.45}$. This is not intuitive
as one could expect the meridional circulation velocity to increases with the rotation rate. A careful study of the $\phi$ component of the vorticity equation shows that the longitudinal vorticity actually decreases with rotation rate as more and more kinetic energy is being transferred to longitudinal motions at the expense of meridional motions.
The fact that in recent 3-D simulations \citep{2007ApJ...669.1190B,2008ApJ...689.1354B,2011AN....332..897M,augustson12}, the meridional circulation is found to weaken as the models is rotated faster implies that standard advection dominated flux transport dynamo models yield the opposite dependency with rotation than the one observed, e.g. activity cycles are found to be longer for faster rotating stars \citep{2010A&A...509A..32J}. This fact alone requires us to revise our current dynamo paradigm for solar-like stars. One way is to short-circuit the advection path by, for instance, adding more cells in latitude or increasing the radial diffusion as was done in \citep{2010A&A...509A..32J,2014ApJ...782...93H}; another is to introduce magnetic turbulent pumping \citep{2008MNRAS.391..467M,2008A&A...485..267G,docao12}. These large-scale magnetic flux transport processes, e.g. meridional circulation, turbulent diffusion and turbulent (or topological) pumping of magnetic flux are extremely important for dynamo operation because they link the spatially segregated source layers of the dynamo, specifically coupling the near-surface layers to the deep convection zone. The competition between these processes influences the overall behavior of the dynamo significantly (\citealt{2007A&A...474..239J,yeates08, 2008A&A...485..267G, karak12, docao12}). Meridional circulation is also thought to critically influence the nature of magnetic butterfly diagrams, including their structure, overlap between cycles and cycle periods (\citealt{charbonneau00, nandy02, 2007A&A...474..239J,nandy11,2014ApJ...782...93H}) and also impacts scaling relationships in stellar activity (\citealt{nandy04,2010A&A...509A..32J}). However, the exact profile of the meridional flow remains unconstrained in stars.  For the Sun, evidence for more than one meridional flow cell are being reported  \citep{2002ApJ...570..855H,2007AN....328.1009M,2013ApJ...774L..29Z} and there is serious debate about its relative role vis-a-vis other flux transport processes.

For instance new stellar dynamo models including magnetic pumping \cite{docao12} yield the following dependency for the cycle period $P_{cyc}$:
\begin{equation}
  P_{cyc} \propto v_0^{-0.40} \gamma_{r0}^{-0.30} \gamma_{\theta 0}^{-0.15} \label{scaling_reference}
\end{equation}
where $\gamma_{r0}$ and $\gamma_{\theta 0}$ represent the amplitude of the radial and latitudinal turbulent pumping terms.
We see that the cycle period dependency on the meridional circulation is reduced and that if we want to recover the observational trends, in particular the shorter cycle period (again, we assume that $v_0 \propto \Omega^{-0.45}$), and assuming that $\gamma_r/\gamma_\theta$ remains constant, the pumping effect should roughly scale as $\Omega_0^2$. Such a scaling may be too extreme and only systematic 3-D numerical simulations will tell us if this is the case or not. Nevertheless, pumping dominated stellar dynamos are a plausible solution to explain stellar observations.

\subsubsection{Stellar Activity and Magnetism: Confronting Mean Field Dynamo Models with Observations}

The general trends in stellar activity observations, in particular the rotation, cycle amplitude, and cycle period relationship is succinctly expressed when confronted with the stellar Rossby number, $R_o$ = $P_{\Omega}/{\tau_{c}}$, the ratio of the rotation period, $P_{\Omega}$ to the convective turn-over-time of the star \citep{1978GApFD...9..241D,1996ApJ...460..848B,montesinos01,2010A&A...509A..32J}. This number can be calculated using a combination of observations with some theoretical ideas about the nature of stellar convection zones, assuming that the convective turn-over-time is $\tau_c = L/V=L^2/\eta$, where $L$ is a characteristic length scale, $V$ a typical convective speed and $\eta$ the magnetic diffusivity. The latter equality is valid if the magnetic Reynolds number $R_m=VL/\eta$ is taken to be of order unity, an assumption that depends on the properties of stellar convection zones and on the convection zone depth. It can be interpreted as characterizing turbulent mixing of magnetic fields in stellar interiors, while the rotation period plays a role in determining differential rotation. It is well known (both theoretically and observationally) that the differential rotation in the convective envelope of solar-type stars is directly connected to the star's rotation rate $\Omega_0$ \citep{1996ApJ...466..384D,2005MNRAS.357L...1B,2007ApJ...669.1190B,2008ApJ...689.1354B,2008JPhCS.118a2029K,2011AN....332..897M,augustson12}. However, the exact scaling $n_r$ (i.e. $\Delta \Omega \propto \Omega_0^{n_r}$) is still a matter of debate among the observers and theoreticians, being sensitive to the observational techniques used and to the modelling approach (see \S 2.2).
Note that the $\alpha$-effect used in mean field theory is related to helical turbulence and represents a measure or parameterization of the electromotive mean force (emf) \citep[e.g.][]{Moffatt:1978tc,1976JFM....77..321P}. It is thus also linked to the rotation rate of the star and the amount of kinetic helicity present in its convective envelope. However in Babcock-Leighton flux transport dynamo models the classical $\alpha$-effect is replaced by a surface term linked to the tilt of the active regions with respect to the east-west direction \citep[the so-called Joy's law, e.g.][]{1919ApJ....49..153H,babcock61,2008ApJ...688L.115K,2010A&A...518A...7D}. This tilt is thought
to be due to the action of the Coriolis force during the rise of the toroidal structures that emerge as active regions \citep{1993A&A...272..621D}. Recent 3-D simulations in spherical shells \citep{2009ApJ...701.1300J,2009LRSP....6....4F,2011ApJ...741...11W,2013ApJ...772...55P,2013ApJ...762....4J,2013SoPh..287..239W} seem to indicate that this is not the only effect responsible for the observed tilt and that the twist and arching of the toroidal structures as well as
the continuous action of the surface convection during the emergence influence the resulting tilt.
The Rossby number, overall, characterizes stellar convection zones -- the zone where the dynamo mechanism generates stellar magnetic fields. Not surprisingly, the Rossby number is simply related through scaling arguments to the dynamo number, ${D}$ ${\sim}$ $1/{R_o}^2$. Thus the Rossby number can be utilized to relate stellar observations to the dynamo mechanism (see e.g., \citealt{noyes84a, noyes84b, tobias98, montesinos01, nandy04}).

On can also implement a non linear feed back via the large scale Lorentz force (also called the Malkus-Proctor effect; \cite{1975JFM....67..417M}) on the dynamo field generation. This has the advantage to yield a saturation (quenching) process based on the actual dynamics of the models \citep[][and references therein]{1997A&A...322.1007T,2000MNRAS.315..521M,2006MNRAS.371..772B}. These models demonstrate that for low magnetic Prandtl number (e.g. $P_m=\nu/\eta$), such that the viscous time scale is much longer than the magnetic diffusion time scale, the dynamo is irregular, with grand minimum-like behavior and switching of parity between symmetric and antisymmetric dynamo families as observed in the Sun \citep{2012ApJ...757...96D}.

From the observational perspective, analysis shows that stars with low Rossby numbers typically exhibit strong activity amplitudes and have irregular cycles, while those stars with relatively higher Rossby numbers tend to have lower activity levels, are more likely to host magnetic cycles and are more likely to be found in grand minima like phases. Since lower Rossby number indicates a higher, underlying dynamo number, and vice-versa, this result is consistent with dynamo  theory. Stars which host very efficient dynamos (having high $D$) are expected to have high activity levels; the non-linear nature of the dynamo mechanism, in conjunction with a high dynamo number also generates irregular cycles because the magnetic feedback is likely stronger. On the other hand, stars with low or moderate dynamo numbers are relatively less active, and are more likely to host regular cyclic behavior with occasional Maunder-like states -- such as the Sun. The consistency of stellar observations with the theoretically expected behaviour of a MHD dynamo mechanism leaves little doubt that internal dynamos are the ultimate source of stellar activity.

A closer look at the activity period versus rotation rate relationship yields more stringent constraints on the nature of stellar dynamos. In particular, it is seen that the periods of stellar magnetic cycles decrease with increasing rotation rate (i.e., decreasing rotation period $P_{\Omega}$). As discussed in \S 2.1, there is an inter-dependance between dynamo action, magnetic activity levels, rotation and stellar evolution via the torque that magnetized thermally driven stellar winds apply on solar-like stars. The mass dependance of this torque is subject to intense studies \citep[see][for a recent discussion]{2013EAS....62..143B}. It is understood that solar dynamo models that rely on the BL mechanism for poloidal field generation are critically dependent on flux transport processes such as meridional circulation which couples the source regions for the toroidal and poloidal components of the magnetic field. Such BL dynamo models have been successful in reproducing various features of the solar cycle. Based on such a BL dynamo model, \cite{nandy04} has shown that the stellar activity amplitude and period versus rotation rate relationship can only be reproduced if both the rotational shear and meridional flow speed scales positively with rotation rate. While full MHD numerical simulations show that the former holds, they also indicate that meridional circulation speed does not increase with rotate rate (\citealt{augustson12}). \cite{docao12} point out in a recent study that the inclusion of turbulent pumping, with the speed of the former increasing with rotation rate, is necessary to recover the stellar scaling laws in the light of these new developments in full MHD simulations. \cite{2013ApJ...775...69D} have also computed 2-D kinematic dynamo using ingredients (such as the alpha effect) deduced from 3-D simulations. They show that essential ingredients of the complex 3-D dynamics (such as for instance the cycle period or the sense of propagation of the dynamo wave)  can be recovered via such techniques. 3-D kinematic dynamo models or convective B-L models have also recently been developed \citep{2007PEPI..163..251C,2012ApJ...746L..26M,2014ApJ...785L...8M}.

Hence confronting stellar activity observations with diverse simulation techniques is very useful and allows the community to test new ideas and concepts.

We now discuss in more details nonlinear, 3-D MHD simulations of solar convection zone and magnetism.

\subsection{3-D Global Models of Stellar Convection and Dynamo Action}

With the advent of massively parallel computers it is now possible to attack the difficult problem of stellar dynamics and magnetism with direct 3-D nonlinear numerical simulations.
As is frequently the case with multidimensional MHD numerical simulations, two approaches are possible: local
high resolution simulations that describe the small scale turbulent motions and local dynamo \citep{1999ApJ...515L..39C,2007A&A...465L..43V}
and global ones that take into account the correct geometry and the existence of
large scale flows in rotating convective zones \cite{1985ApJ...291..300G,1985GApFD..31..137G,2004ApJ...614.1073B,2009AnRFM..41..317M}.
Given the broadness of the topic, here we present the progress made in modeling
stellar convection and dynamos with the global approach. These simulations allow us to describe turbulent, rotating and magnetized convection either in a full sphere, a shell or a wedge-like geometry \citep[cf. the recent book by][]{2013SASS...39.....C}.
Contrary to the kinematic mean field stellar dynamo approach discussed in the previous sections, 3-D simulations use the
full set of MHD equations \citep[we refer to these recent references for simpler 2-D models of stellar differential rotation assuming a parametrization of turbulent convection][]{2008JPhCS.118a2029K,2011ApJ...740...12H,2012MNRAS.423.3344K}.
These equations couple classical fluid mechanic equations to the induction equation and
as such do not prescribe the velocity field but compute it self consistently \citep[c.f.][]{2004ApJ...614.1073B,2005LRSP....2....1M}. This is the main advantage
of the global full MHD approach: the influence of rotation and magnetic field on turbulence and convection is
computed not parametrized.

Over the last three decades several groups have developed stellar convection and dynamo models in
spherical geometry following the pioneering work of P. Gilman and G. Glatzmaier in the early-mid 80's \citep{1981ApJS...46..211G,1982ApJ...256..316G,1983ApJS...53..243G,1985ApJ...291..300G,1985GApFD..31..137G}.
Prominent examples of modern global or wedge-like parallel codes that have in recent years published original studies on stellar convection and dynamo include:
\begin{itemize}
\item the Anelastic Spherical Harmonics (ASH) code: \cite{2000ApJ...532..593M,2002ApJ...570..865B,2007ApJ...669.1190B,2008ApJ...673..557M,2008ApJ...676.1262B,2011ApJ...728..115B,augustson12} on surface convection and zonal flows and \cite{2004SoPh..220..333B,2004ApJ...614.1073B,2006ApJ...648L.157B,2008ApJ...689.1354B,2010ApJ...711..424B,2011ApJ...731...69B,2011ApJ...742...79B,2012ApJ...746L..26M,2013ApJ...762...73N,2013ApJ...777..153A,2014SoPh..289..441N} on solar-like star dynamo action, cycles and magnetic wreaths,
\item the Eulag MHD code: \cite{2010ApJ...715L.133G,2011ApJ...735...46R,2013SASS...39..187C} on solar and stellar dynamo and cycles,
\item the PENCIL code: \cite{2013ApJ...778..141W,2013ApJ...778...41K,2013ApJ...779..176G} on stellar convection, dynamos and cycles,
\item the Magics code: \cite{2006GeoJI.166...97C,2009Natur.457..167C,2010SSRv..152..565C,2013Icar..225..156G,2014MNRAS.438L..76G} on convection and dynamos
\item K. Chan's code \cite{2011JCoPh.230.8698C} on core convection
\end{itemize}
Other codes have been mostly used to model the geodynamo with some applications to the dynamo in M dwarfs.
\begin{itemize}
\item Glatzmaier's code: adapted to model Earth's dynamo \cite{1996Sci...274.1887G} or giant planets dynamics \cite{2009GApFD.103...31G},
\item EBE code: \cite{2007PEPI..163..251C},
\item Leeds code \cite{1993Natur.365..541H,2006GeoJI.164..467S} ; dynamo review by \cite{2011AnRFM..43..583J}
\item PARODY code: \cite{2008EL.....8359001G,2011MNRAS.418L.133M,2012ApJ...752..121S,2014A&A...564A..78S} on geo and stellar dynamos and \cite{2007PEPI..160..143A} on the role of thermal wind in setting mean large scale flows and \cite{2008GeoJI.172..945A} on dynamo
\item Busse and Simitev's code: \cite{2001FlDyR..28..349G,2009EL.....8519001S} on dynamo and bistability
\item Aurnou and Heimpel's code: \cite{2005E&PSL.236..542H,2007Icar..190..110A} on dynamo in convective envelopes
\item Several research groups in Japan: \cite{1999GApFD..90...43T,2004GeoRL..3112609Y,2011PhFl...23g4101K} on convection or zonal flows or \cite{1997PhRvE..55.4617K,2008PhPl...15h2903N,2013ApJ...778...11M} on dynamo, solar convection and magnetism, the most recent studies making use of a novel ``ying-yang'' grid to map a spherical shell \citep{2004GGG.....5.9005K}.
\end{itemize}

\noindent Many of the codes used in the geophysics community were originally developed to model the geodynamo.  They therefore rely on the Boussinesq approximation,
which is not appropriate for highly stratified and compressible convective envelope as found in solar-like stars. Modelers have been converting these codes to be able
to deal with stellar-like stratified flow by using the anelastic approximation \citep{1969JAtS...26..448G,1984JCoPh..55..461G,2013ApJ...773..169V} and validating their codes
with a recent international benchmark \citep{2011Icar..216..120J}.

Given the intrinsic complexity of computing nonlinear convection and dynamo simulations and the large diversity of stars to model,
we first briefly describe purely hydrodynamical models (see also chapter 3 in the ISSI book \citep{sismoissi}). 
Such models reveal the fundamental ingredients at the origin of large scale mean flow, such as differential rotation 
and meridional circulation in rotating global convection models. 
We then follow up by describing the origin of magnetic fields (either cyclic or irregular) through  nonlinear dynamo action and the feedback 
of such fields on the flows -- making the link with the 2-D models discussed earlier when necessary.

\subsubsection{Convection and Large Scale Flows}

Convection is present in stars of all masses on the main sequence. It takes either the form of a convective core in massive ones (greater than about 1.3 $M_{\odot}$) or
of a convective envelope of varying thickness for lows mass stars (from early F down to about M3), with stars less massive than about 0.3 $M_{\odot}$ being fully convective.
Here we will limit our discussion to solar-like stars possessing a deep convective envelope. For a discussion on two other categories (e.g. high mass and very low mass stars) interested readers may refer to  \cite[e.g.][and references therein]{2004ApJ...601..512B,2005ApJ...629..461B,2006ApJ...638..336D,2008ApJ...689.1354B,2009Natur.457..167C,2009ApJ...705.1000F,2011IAUS..271..361A,2011JCoPh.230.8698C,2011MNRAS.418L.133M,2012ApJ...752..121S}.

\begin{figure}[!htb]
\begin{center}
\includegraphics[width=0.45\textwidth]{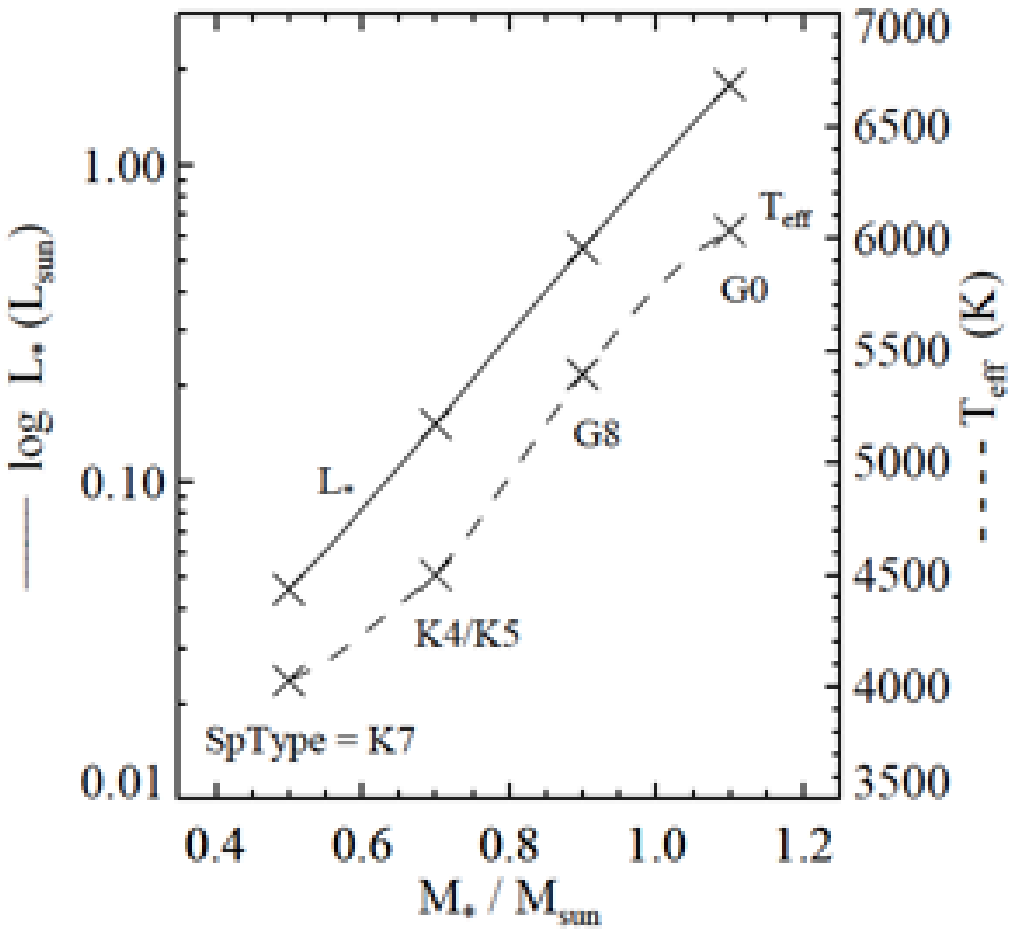}
\includegraphics[width=0.45\textwidth]{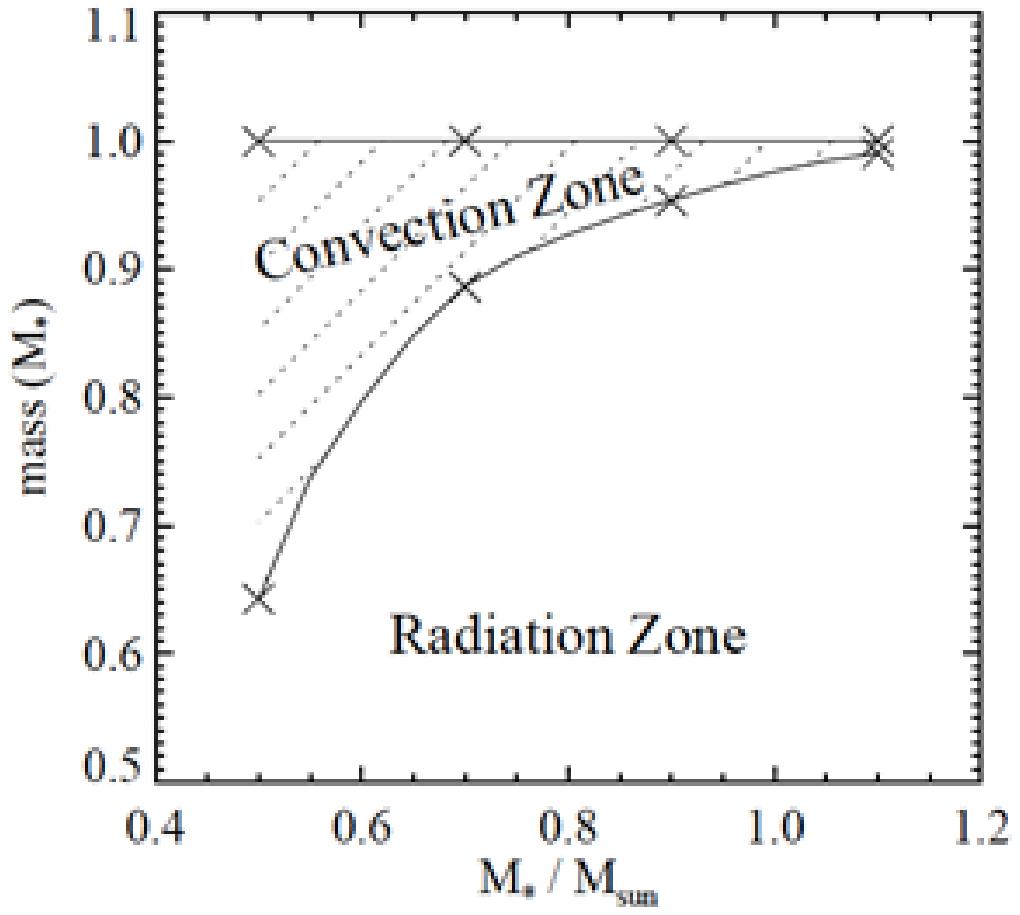}
\end{center}
\caption{Stellar luminosity, effective temperature and mass contained in the convective envelope vs spectral type in solar like stars with
the solar metallicity \citep[computed with the CESAM code][]{1997A&AS..124..597M}}
\end{figure}

\begin{figure}[!htb]
\begin{center}
\includegraphics[width=0.9\textwidth]{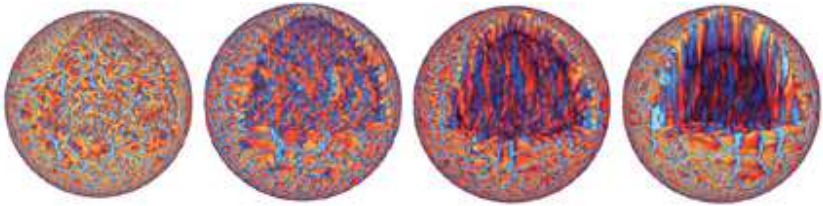}
\end{center}
\caption{\label{conv} 3-D rendering of turbulent surface convection in solar-like stars with varying convection depths \citep{2011ApJ...728..115B}.
The radial velocity is shown with cold downflows in blue tones.}
\end{figure}

Convection plays a key role in transporting the star's heat by having strong correlations between vertical motions and temperature fluctuations.
It is a fundamentally turbulent process in stars as the Rayleigh number ($Ra$, e.g. the ratio between buoyancy driving and dissipation process) is
huge, of order $10^{18}$ in the Sun. This leads to a large range of scales from possibly giant cells on the order of the thickness
of the convective layer (e..g 200 Mm in the Sun) down to granulation size (1 $Mm$), with intermediate scales such
as supergranulation (20-50 Mm) well delineated by the surface magnetic network \citep{2010LRSP....7....2R}. As the luminosity (energy output) of
the star changes so does the amplitude of the convective flow. A simple scaling argument based on mixing length treatment (MLT) gives
$v_r \propto (L_*/(\rho_{cz} R_*^2))^{1/3}$, with $L_*$ and $R_*$ respectively the star's luminosity and radius and $\rho_{cz}$ the averaged density of the convective envelope  \citep{2013sse..book.....K}. Since the luminosity varies by at least 2 orders of magnitude across the spectral type we are considering (i.e early F to early M) ($L_* \propto M_*^4$), the stellar radius only by a few ($R_* \propto M_*^{0.9}$), it is expected that the vigor of the convective flows do likewise modulo the effect of density. As can been seen in Figure 8, 1-D stellar structure models of solar-like stars indicate that the averaged mean density changes significantly in the convective envelope of these stars, in such a way that more massive stars have on average a lower mean density than low mass ones \citep{2011AN....332..897M}. This is easily understood as the base of the envelope moves up towards the surface with the thinning of the convective envelope. This yields a large amplitude difference between F and K stars for their characteristic convection velocity.
This difference in velocity amplitude is recovered in global numerical simulations of stellar convection. 
We illustrate an example of such 3-D turbulent convective simulations in Figure \ref{conv}, where we clearly see the complex network of downflows
surrounding broad upflows.

The turbulent motions within the convective envelopes of rotating stars experience the effect of the Coriolis force to a degree that depends on their fluid Rossby number $Ro=\omega/2\Omega_*$, here defined as the ratio between turbulent and the so-called planetary vorticity \citep{1982bsv..book.....P}.
We note that several definitions of the Rossby number exist in the community (stellar Rossby based on the ratio between convective turnover time and stellar rotation period; convective Rossby$\sqrt{Ra/TaPr}$; local Rossby number), but it can be shown that they depend almost linearly on one another \citep[see for example][]{1996ApJ...457..340K,2010A&A...510A..46L,matt14}. We here choose to use the fluid Rossby $Ro$ because a clear transition of behavior occurs at $Ro \sim 1$ with the differential rotation becoming prograde or retrograde. Indeed if the stars rotate fast enough such that the large scale motions (larger than the Rossby radius of deformation) feel the continuous action of the
Coriolis force  as they rise and sink then the redistribution of angular momentum is such that a non uniform rotation profile is established in the convective envelope.

\begin{figure}[!htb]
\begin{center}
\includegraphics[width=0.2\textwidth]{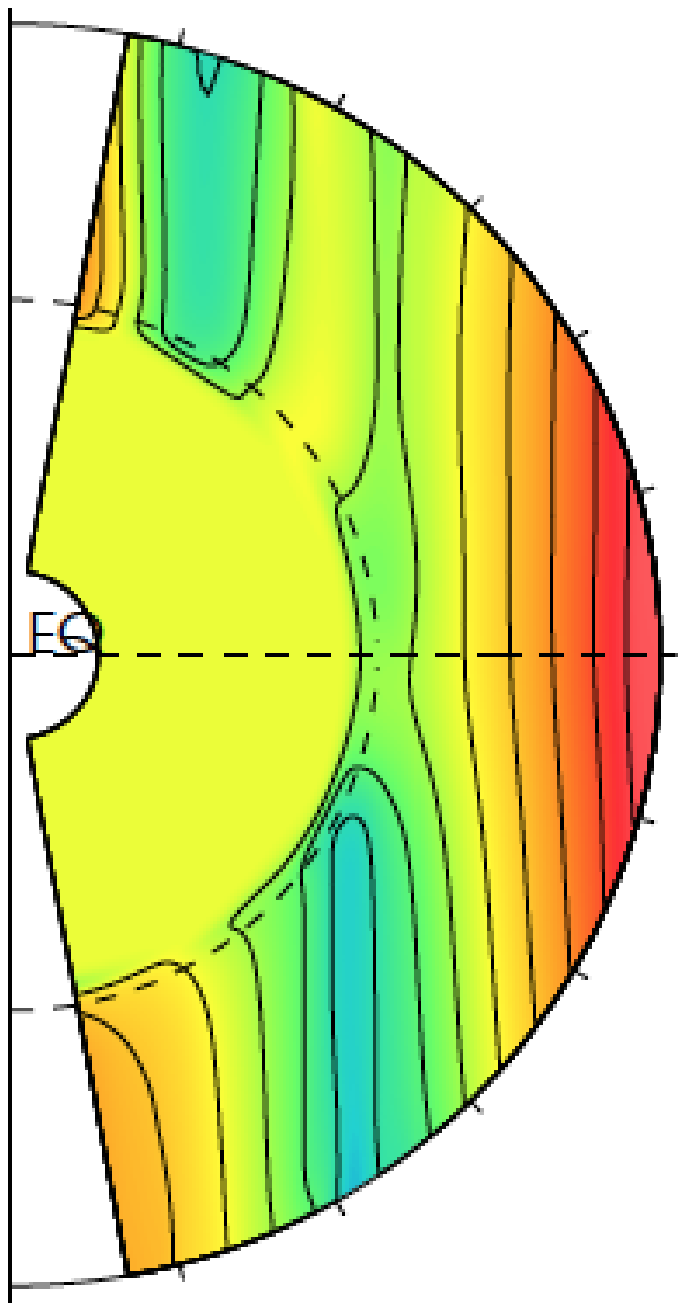}
\includegraphics[width=0.2\textwidth]{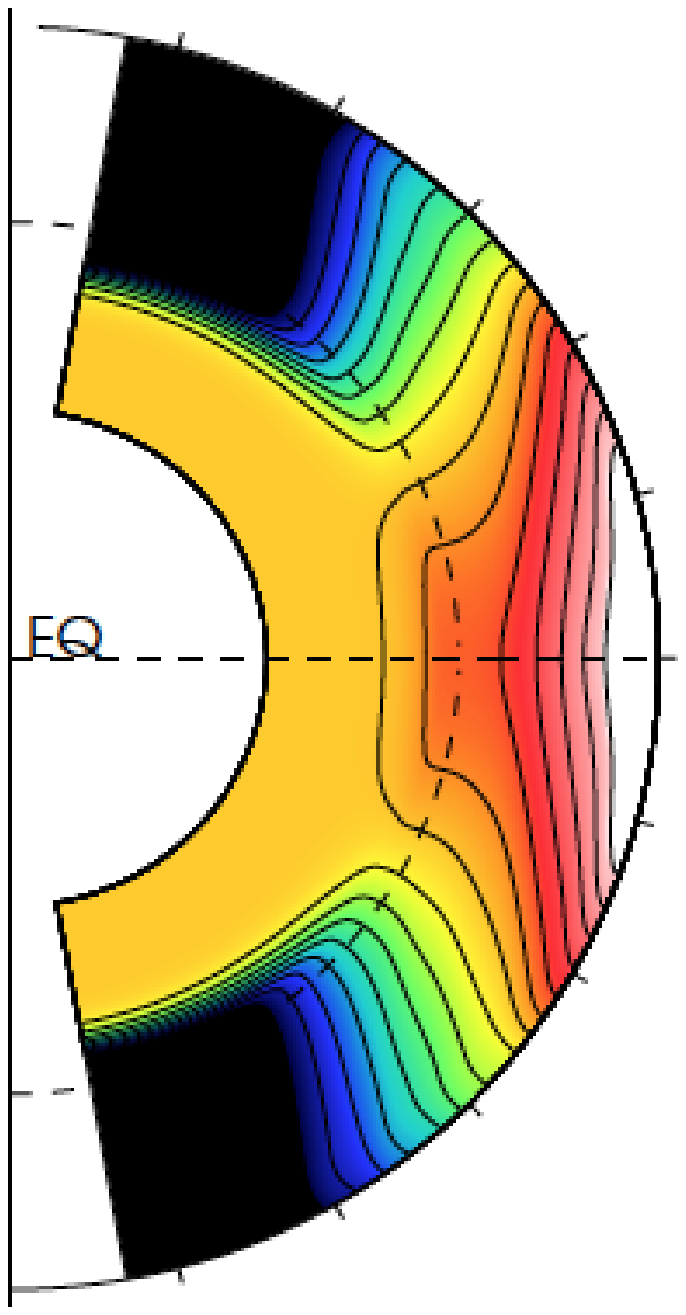}
\includegraphics[width=0.2\textwidth]{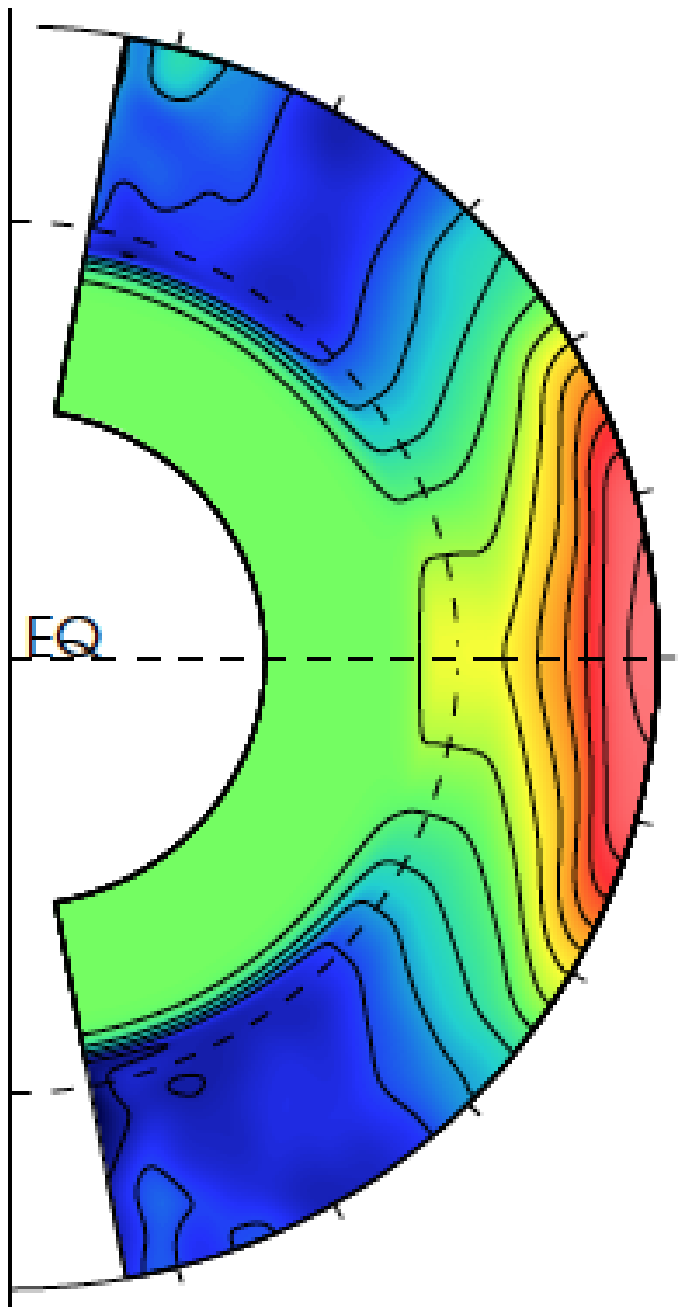}
\includegraphics[width=0.2\textwidth]{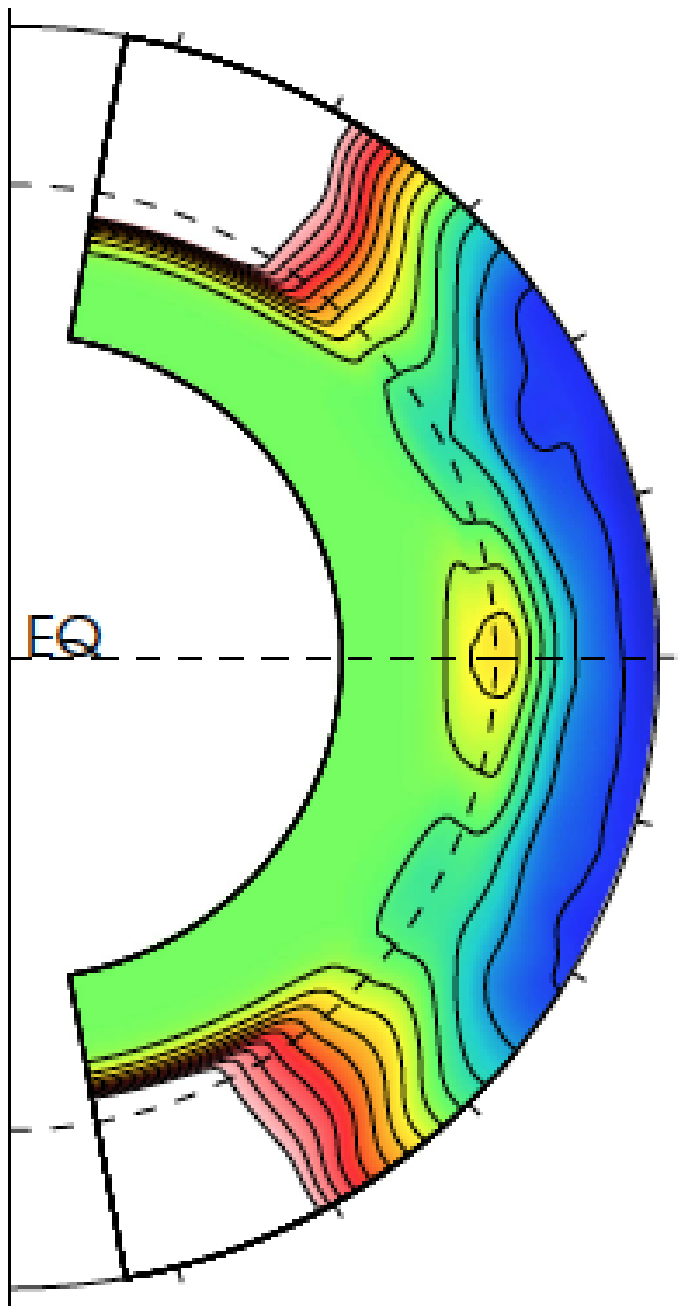}
\includegraphics[width=0.2\textwidth]{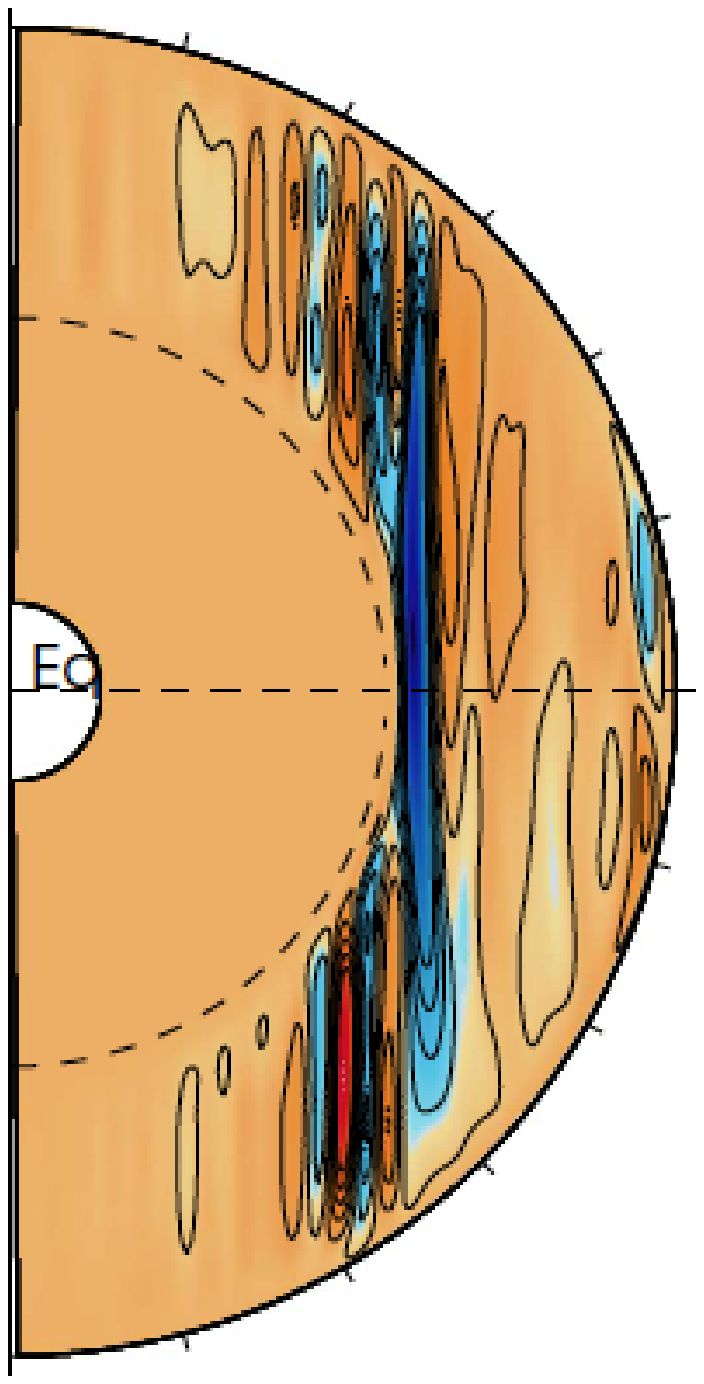}
\includegraphics[width=0.2\textwidth]{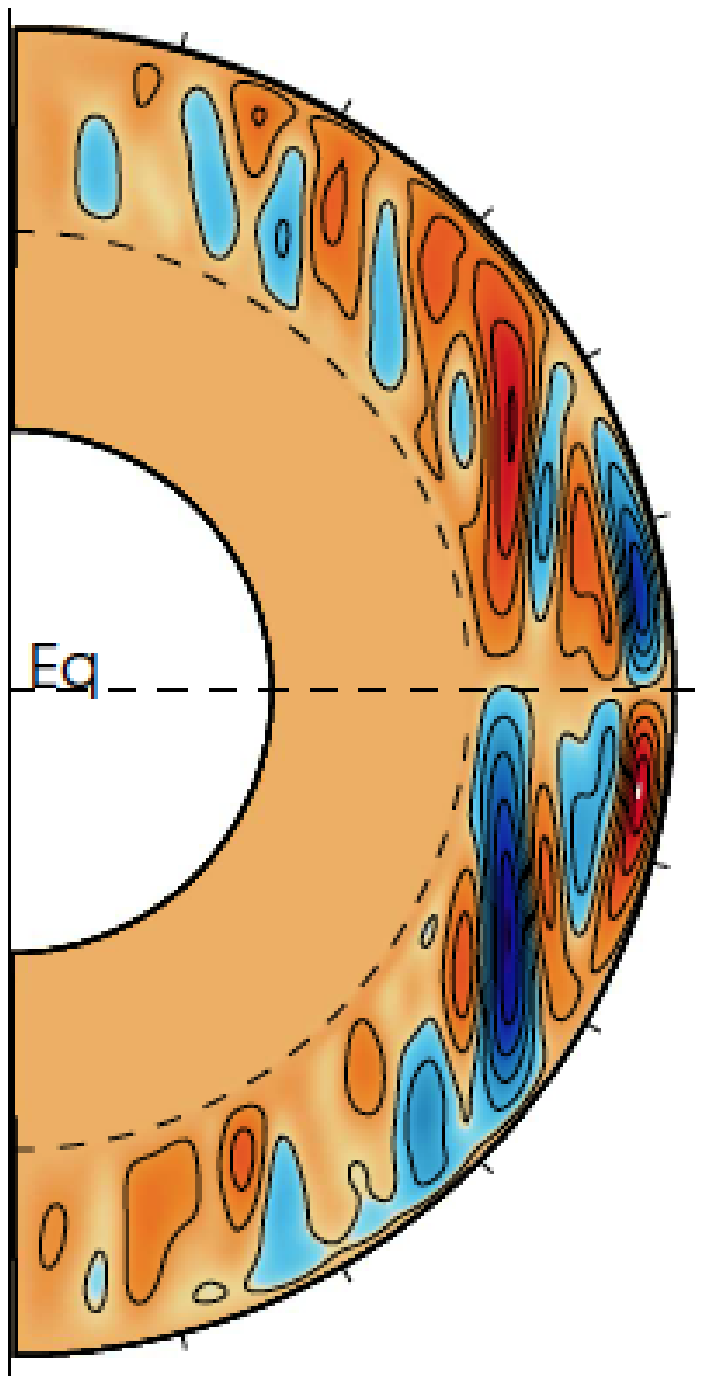}
\includegraphics[width=0.2\textwidth]{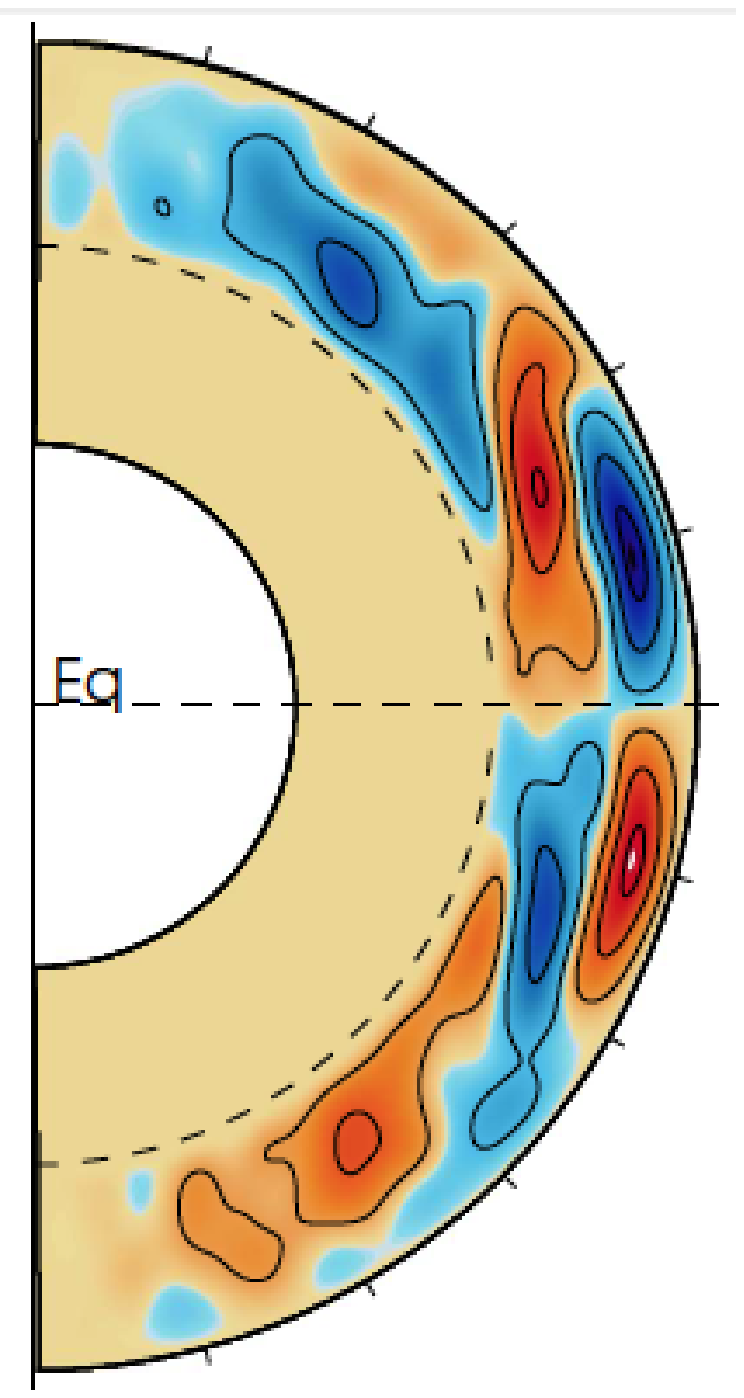}
\includegraphics[width=0.2\textwidth]{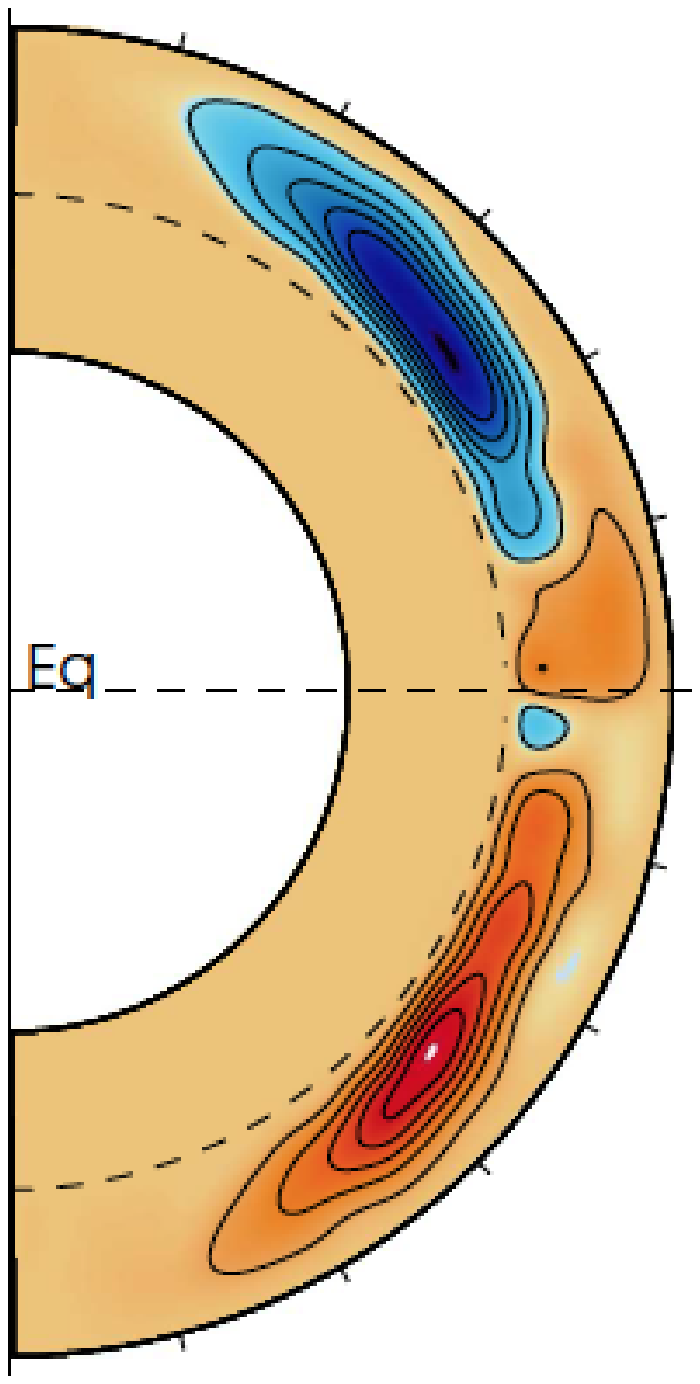}
\end{center}
\caption{Differential rotation and meridional circulation realized in 3-D models of
solar-like convective envelope coupled nonlinearly (i.e both zones back react on one another) to a stably stratified interior \citep[][in preparation]{matt14}).
3 masses (0.5, 0.9 and 1.1) $M_{\odot}$ and 3 rotation rates (solar, 3 and 5 time solar) are being shown.
Starting from the left in each row: color contours showing $\Omega(r,\theta)$ and the meridional streamfunction for a
0.5 $M_{\odot}$ star rotating at 5 $\Omega_{\odot}$, 0.9 $M_{\odot}$ star rotating at 3 and 1 $\Omega_{\odot}$ and 1.1 $M_{\odot}$ star rotating at $\Omega_{\odot}$.
Prograde rotation is shown in red/white colors and retrograde rotation in dark/blue. Clockwise meridional flows in red.}
\label{meanflow}
\end{figure}

In figure \ref{meanflow} we display 4 realizations of the longitudinally averaged differential rotation and meridional circulation computed with the ASH code \citep[][in preparation]{matt14}. These profiles have been obtained by changing either the stellar mass (i.e aspect ratio and luminosity) or the reference frame rotation rate. We note several trends: As the rotation rate is increased, the differential rotation becomes more and more aligned with the rotation axis (along $\hat{e}_z$) and likewise the meridional circulation posses more and more cells both in latitude and radius. The 1.1 solar mass case rotating at one time solar rate exhibits an anti-solar differential rotation, with slow equator and fast pole. It is indeed the only case with a Rossby number greater than 1 as seen in many recent studies \citep{2011ApJ...728..115B,2011AN....332..897M,2013ApJ...779..176G,2014MNRAS.438L..76G,2014arXiv1401.2981K}. We thus see that  the transition between prograde and retrograde differential rotation will not occur at the same rotation rate ($v \sin i$) for each stellar spectral type. We also note the strong shear layer at the base of the convective domain \citep[e.g. the tachocline][]{1992A&A...265..106S} that naturally develops in each modeled stars with a profile closer to the solar case for all the cases with a Rossby number $Ro \sim 0.6$.

These profiles can be understood by considering the transport of angular momentum in a spherical convective shell. When performing this analysis as in \citep{2002ApJ...570..865B,2008ApJ...689.1354B,augustson12}, we find that the Reynolds stresses are at the origin of the prograde equator for models with a low Rossby number. Through gyroscopic pumping the meridional circulation transport angular momentum to counterbalance the torque applied by the Reynolds stresses (from which we have subtracted the contribution from viscous effect). We refer to \citep{1992A&A...265..115Z,2004A&A...425..229M,2007sota.conf..183M,2010ApJ...719..313G,2011ApJ...742...79B,2011ApJ...743...79M} for a more thorough discussion of the role of the meridional circulation in reaching angular momentum balance in rotating stars. The orientation of the iso-contours of $\Omega$ from being cylindrical to being more conical (solar-like) at mid latitude is due to the effect of an efficient thermal wind associated with latitudinal entropy variations (with relatively hotter poles and cooler equator) \cite{2006ApJ...641..618M}.
These variations are enhanced further by the presence of a tachocline \citep{2011ApJ...742...79B}.

\begin{figure}[!htb]
\begin{center}
\includegraphics[width=0.45\textwidth]{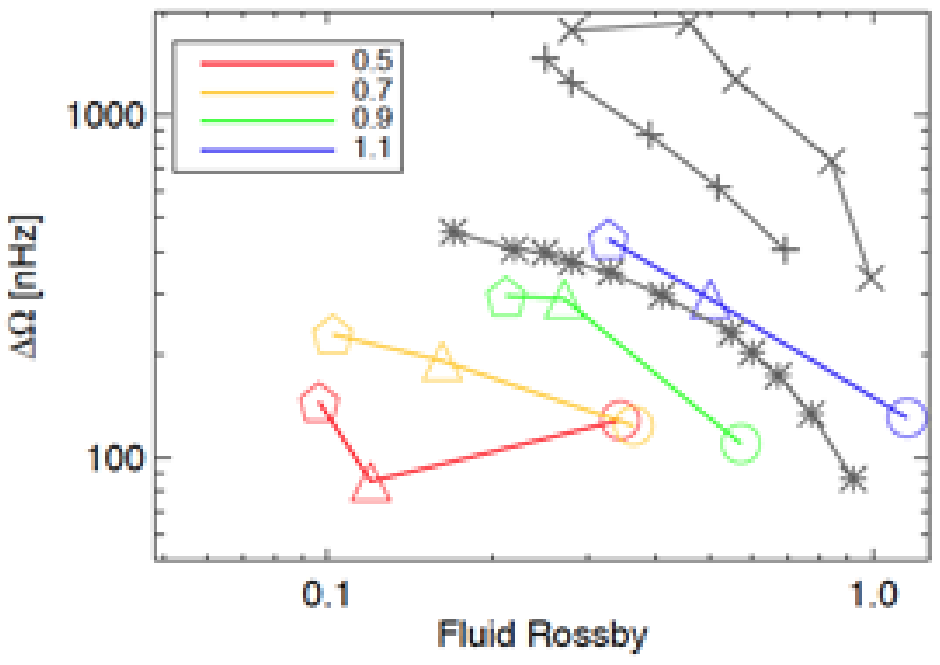}
\includegraphics[width=0.48\textwidth]{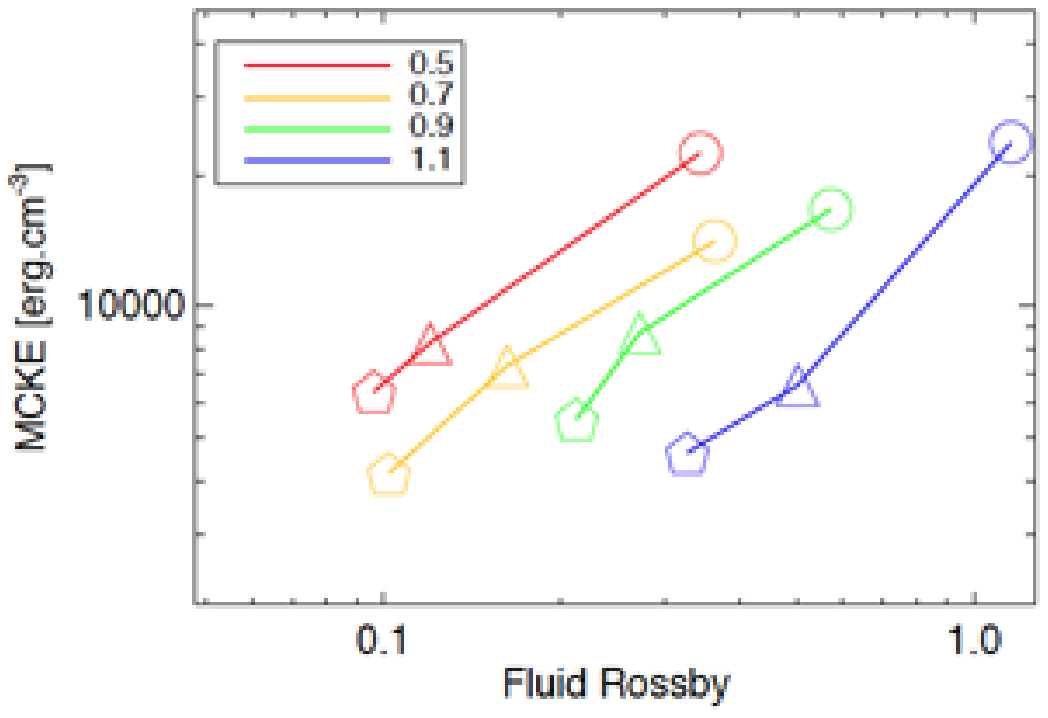}
\end{center}
\caption{Differential rotation $\Delta \Omega$ as a function of Rossby number for stellar masses ranging from 0.5 up to 1.3 $M_{\odot}$.
1 $M_{\odot}$ cases from \citep{2008ApJ...689.1354B}, 1.2 and 1.3 $M_{\odot}$ cases coming from \citep{augustson12}).
Kinetic energy of the axisymmetric meridional circulation as a function of the Rossby number for models ranging from 0.5 to 1.1 $M_{\odot}$, from \citep[][in preparation]{matt14}.}
\label{scaling}
\end{figure}

One important aspect of performing 3-D numerical simulations is that one can obtain general trends on the variations of the main physical processes acting in convective envelope with various parameters such as rotation rate, luminosity, aspect ratio. We focus on scaling laws for the differential rotation and meridional circulation by synthesizing the recently published 3-D stellar convection models computed with the ASH code. The models range from 0.5 up to 1.3 times the solar mass and from 1 to 20 times the solar rotation rate \citep{2008ApJ...689.1354B,2011AN....332..897M,augustson12}.
In Figure \ref{scaling} we show how the latitudinal differential rotation contrast varies as a function of the Rossby number. There is a clear trend for $\Delta \Omega$ to rise with increasing rotation rate.
This is in qualitative agreement with direct and indirect observational data that imply larger absolute shear rates in rapid rotators to produce stronger magnetic fields and enhanced angular momentum loss relative to slow rotators. The actual slope seems a bit larger than observed by \cite{2007AN....328.1030C} but not so when compared with the work by \cite{1996ApJ...466..384D}. \cite{2011IAUS..273...61S} found that the presence of binaries in the published samples, which can be distinct from single stars because of tidal effects have led to an underestimate of the dependency of $\Delta \Omega$ on rotation rate. Clearly more observational data are required to constrain better the exponent $n$ in the equation $\Delta \Omega \propto \Omega^n_*$. We also clearly see that more massive stars possess larger $\Delta \Omega$ for a given Rossby number. This is in reasonable agreement with the observations published in \cite{2005MNRAS.357L...1B}. Although, given the paucity of data and the rather large error bars, the exact exponent is uncertain. Data from the \emph{Kepler} mission and in the (future) PLATO mission will be a major asset in this context \citep{2014arXiv1403.7155G,2013arXiv1310.0696R}.

Another interesting trait is that the meridional circulation is found to decrease in strength as the stars are made to rotate faster. Rapid rotation causes more kinetic energy to go into toroidal motions, which can be anticipated by looking at the equation for gyroscopic pumping \citep{2010A&A...510A..46L,augustson12}.
This has a very important consequence: Flux-transport dynamo models as discussed in the previous section can't be easily extended to other stellar spectral type to explain the dependence of their cycle period with their rotation period. Indeed flux-transport models of the advective type depends directly on the meridional circulation amplitude to set the cycle period. Observations indicate that the cycle period should decrease with increasing rotation rate. However, given that the meridional flow is found to weaken, this can only result in an increase of the activity cycle as was shown by \cite{2010A&A...509A..32J}. One needs either to take into account multi-cellular flows as observed also in 3-D simulations \citep{2010A&A...509A..32J} or take into account turbulent pumping \citep{docao12}.

\subsubsection{Nonlinear Dynamo Effect, Magnetic Activity, and Cycles}

The observed magnetic activity shows the existence and period of cycles seem to depend on stellar spectral type and rotation rate. An obvious trend is that old (older than the age of the Hyades cluster $\sim 800\, Myr$), slowly rotating solar-like stars are less active than their younger counterparts. We have also discussed in the previous section that convection and large scale flows change with spectral type. It is certainly not surprising that the associated dynamo action and magnetic activity do so as well.

Several groups have recently published extensive parameter studies of dynamo action in rotating convective spherical shells of various thickness and degree of stratifications. The key results can be summarized as follow:
3-D numerical simulations of dynamo action in solar-like stars have revealed a large range of behavior, from steady dynamo, to irregular and cyclic ones \citep[][and references therein]{2004ApJ...614.1073B,2010ApJ...711..424B,2011ApJ...731...69B,2011ApJ...735...46R,2012A&A...546A..19G,2012ApJ...752..121S,2013ApJ...777..153A,2013ApJ...778...41K,2013ApJ...762...73N,2014A&A...564A..78S}.

For low mass (M-type) stars, it is seen that a bistability similar to that found for the geodynamo is possible. Observations by \cite{2010MNRAS.407.2269M} seems to indicate that a strong dipolar state
and a multipolar state may coexist for stars having similar stellar parameters. Dynamo models show that around a local Rossby number of 0.1 a weak (multipolar) and strong (dipolar) dynamo branch may coexist \cite{2006GeoJI.166...97C,2011MNRAS.418L.133M}. Such bistable states are found in a given range of parameters (low stratification, high Pm) and may not actually exist in real stars. It may be the case that these stars are observed in a different phase (low vs high) of their activity cycle. Other similarity between the geodynamo and low mass stellar dynamos also exist such as the scaling of the magnetic energy with the available heat flux \citep{2009Natur.457..167C}.
Such scaling tends to break down for higher mass stars for which the tachocline may play a significant role.

\begin{figure}[!ht]
  \centering
  \includegraphics[width=10cm]{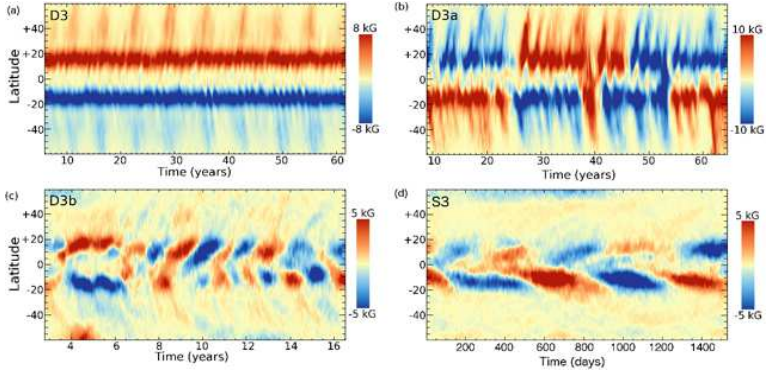}
  \caption{Magnetic wreaths yielding in turn steady (case D3; top left), irregular (case D3a; top right) and quasi cyclic (cases D3b \& S3; bottom left and right) magnetic butterfly diagrams \cite{2013ApJ...762...73N}.
  Red tones correspond to positive toroidal polarity.}
  \label{btfly}
\end{figure}

Others 3-D simulations of F, G and K stars show that for high rotation rates, large scale magnetic wreaths (see Figure \ref{bpolrev} left panel) are obtained without requiring the presence of a tachocline \cite{2010ApJ...711..424B,2011ApJ...731...69B}. We show on Figure \ref{btfly}, four butterfly diagrams (time-latitude plots of the azimuthally averaged toroidal magnetic field near the base of the CZ) realized in such simulations for of a solar-like star rotating at three times the solar rate \cite{2013ApJ...762...73N}. We remark that as the model is made more turbulent, the steady magnetic wreaths become more time dependent and can lead to cyclic activity (bottom right panel).

In such stars, along with the degree of turbulence, rotation plays an important role in determining the global properties of their magnetism. This is due
to a shift in the balance of forces driving the flow between the advection, Coriolis and Lorentz terms. As the rotation rate increases the Lorentz
force tends to balance the Coriolis force yielding larger magnetic energy in superequipartion with the kinetic energy of the flow (a direct consequence of
a magnetostrophic state; c.f. strong scaling below) as in the Earth's iron core. The Elsasser number $\Lambda=B^2/4\pi\bar{\rho}_{cz}\eta\Omega_0$, with $\bar{\rho}_{cz}$ mean density in the convective envelope, $\eta$ magnetic diffusivity, $\Omega_0$ stellar rotation rate, $B$
a characteristic magnetic field of the CZ, is useful to discuss this balance of terms in the Navier-Stokes (N.V.) equation.
Depending on the amplitude of this number and on the balance assumed in the Navier-Stokes equation, various scaling of the magnetic field amplitude can be expected \citep{2007CRPhy...8...87F,2010SSRv..152..565C}:

\begin{figure}[!htb]
\begin{center}
\includegraphics[width=0.34\textwidth]{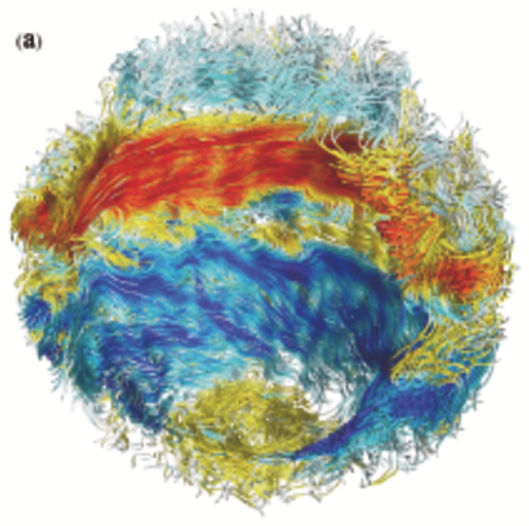}
\includegraphics[width=0.61\textwidth]{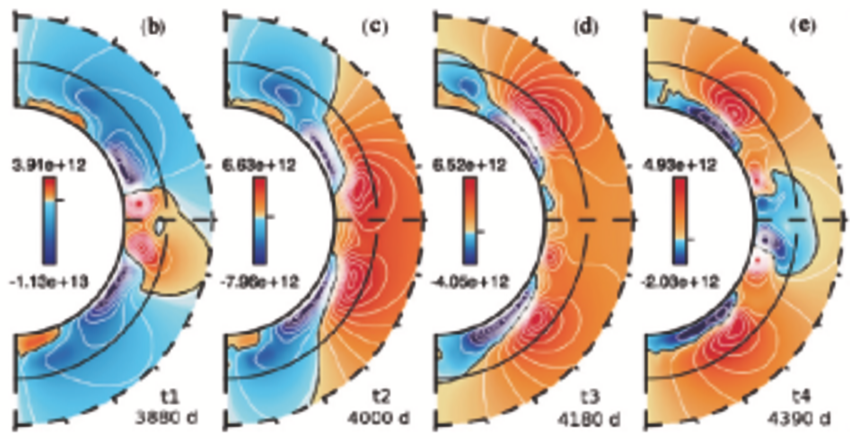}
\end{center}
\caption{Magnetic wreaths shown by 3-D rendering of magnetic field lines \citep{2010ApJ...711..424B}. Temporal sequence of meridional cuts showing the mean poloidal field realized in
case D5 of \cite{2011ApJ...731...69B} which was extrpolated using potentiel field approximation to about 1.5 stellar radius. Note the global polarity field reversal of the poloidal field.}
\label{bpolrev}
\end{figure}

\begin{itemize}
\item To an order of magnitude, an estimate of the equilibrium magnetic field (assuming ideal gas law) can easily be obtained:
$B_{eq} \sim \sqrt{ 8\pi P_{gas}} \sim \sqrt{\bar{\rho}_{cz}}$, since ${T}_{eff}$ varies by a factor
2 to 3 between early F and late K stars, whereas $\bar{\rho}_{cz}$ varies by more than a factor of 100 (\cite{2011AN....332..897M}).
\end{itemize}

If one assumes for the sake of simplicity that the magnetic Reynolds number $R_m = v d/\eta \sim$1 such that a characteristic velocity is given by $v \sim \eta/d$
\footnote{note however that such choice of characteristic velocity scale is subject to caution in stellar turbulent convective envelopes that usually possess very large $Rm$}, the balance of terms in Navier-Stokes Equation is then given by:

\begin{itemize}
\item  Laminar (weak) scaling: Lorentz $\sim$ viscous diffusion\\
$\mbox{   } \mbox{   } \mbox{   } \mbox{   } \mbox{   } \mbox{   }  \Rightarrow B^2_{weak} \simÊ \bar{\rho}_{cz} \nu v/d \sim \bar{\rho}_{cz} \nu \eta/d^2$

\item Turbulent (equipartition) scaling: Lorentz $\sim$ advection\\
$\mbox{   } \mbox{   } \mbox{   } \mbox{   } \mbox{   } \mbox{   } \Rightarrow B^2_{turb} \simÊ\bar{\rho}_{cz} v^2 \sim \bar{\rho}_{cz} \eta^2/d^2 \Leftrightarrow |B_{weak}| \sim |B_{turb}| P_m^{1/2}$

\item Magnetostrophic (strong) scaling (e.g. Elsasser nb $\Lambda \sim 1$): Lorentz $\sim$ Coriolis\\
$\mbox{   } \mbox{   } \mbox{   } \mbox{   } \mbox{   } \mbox{   } \Rightarrow B^2_{strong} \simÊ\bar{\rho}_{cz} \Omega_0 \eta$
\end{itemize}
with $v$, $d$ are characteristic velocity and length scales, $P_m = \nu/\eta$ the magnetic Prandtl number. Of course there is an upper limit to the magnitude of the magnetic energy
ultimately set by the amount of energy (likely the star's outward energy flux) that can be made available to the dynamo process \citep{1996Natur.384..544S,2009Natur.457..167C}.
We recall here that dynamo action does not exist for all fluid motions due to its intrinsic 3-D character \citep{Moffatt:1978tc}.

So a possible scenario for stellar dynamos based on heuristic scaling arguments could be the following:
Stars rotating at moderate rate (such that their Elsasser number is small), have a level of magnetic energy (or averaged global field strength) that are less than or of the order of the equipartition field given by either the weak or turbulent scalings described above. As stars rotate faster and get closer to a magnetostrophic state with an Elssasser number of order 1 or larger, the formation of large and intense magnetic wreaths starts. The magnetic field becomes more and more dominated by its toroidal component and the magnetic energy becomes larger and larger going above the equipartition value set by the turbulent scaling as it now follows the strong scaling which depends linearly on the rotation rate. The consequence may be the following. As the magnetic energy (or field amplitude) becomes large, the associated Lorentz force starts back reacting strongly on the mean flow. The first consequence is what can be called an ``$\Omega$-quenching'', e.g. the differential rotation reduces in strength and an almost solid body rotation state in the convective zone (envelope or core) is established \citep{2004SoPh..220..333B,2005ApJ...629..461B}. Using mean-field classification, one may then hypothesize that the stellar dynamo transitions from an $\alpha-\Omega$ or $\alpha^2-\Omega$ to being an $\alpha^2$ dynamo, i.e. helical turbulence is solely responsible for field generation and maintenance, the large scale shear now plays a more marginal role.
At that stage what remains of the magnetic wreaths is still unclear, more work must be done. As the rotation is made even faster, quenching of the $\alpha$ effect, due to the large scale magnetic field being more and more intense, occurs naturally as now the Lorentz force back-reacts directly on the helical convective motions and not solely on the large-scale mean flow. Following this discussion, one question naturally crops up which requires detailed investigations: How are these diverse set of plausible dynamo scaling relationships (involving magnetic activity amplitudes) related to stellar activity trends and can the latter impose more stringent constraints on solar and stellar dynamo processes?

As we have seen in section \S 1.6, the activity of solar-like stars can be assessed by observing their X-ray luminosity $L_x$. Using such proxy it has been shown that magnetic activity increases with rotation rate and that above a given rotation rate (that depends on the stellar spectral type but is around $R_{os} \sim 0.1$), stellar magnetic activity first
saturates \citep{2003A&A...397..147P,2011ApJ...743...48W} and then possibly ``over-saturates'' for extremely fast rotation rate such as the ones found in young solar-like T-Tauri stars \citep{1999ARA&A..37..363F}. One can thus simply seek to characterize how the saturation of the dynamo generated field can be linked to the saturation of solar-like star X-ray emissivity. Such a link is not as straightforward to deduce as one could anticipate. Indeed one must also assess how the filling factor $f$
of the magnetic field on the star's surface evolves with stellar parameters not just the field strength.
As we have noted the field amplitude does not easily saturate since the Lorentz force can achieve a dynamical balance with various terms whose importance and amplitude vary with stellar parameters.
One may conjecture that the first saturation of the X-ray luminosity is due to $\Omega$-quenching and to the physical limitation of the spot coverage (e.g., filling factor $f$) on the stellar surface and
that the second ``over-saturation'' may be due to ``$\alpha$-quenching'', so of the actual field strength (see \citealt{2012A&A...546A.117G} for an alternative explanation).
Currently no global-scale 3-D nonlinear MHD dynamos actually include spot formation and emergence. Global flux emergence simulations have been computed to assess the role of the
Coriolis force, Hoop stresses, convective motions, large scale mean flows and background magnetic fields on the evolution and morphology of simplified active regions by varying the properties
of idealized global flux ropes \citep{2009LRSP....6....4F,2009ApJ...701.1300J,2011ApJ...741...11W,2011A&A...528A.135I,2013ApJ...762....4J,2013ApJ...772...55P,2013SoPh..287..239W} but until very recently, none had self-consistently dealt with the formation of sunspots in a cyclic convective dynamo. So in order to be able to physically characterize the saturation of $B$ and $f$ independently,
{\it spot-dynamos} -- wherein dynamos self consistently produce rising omega-loops -- must be developed and the parameter space explored systematically.

There has been recent progress in this direction using improved numerical treatment in 3-D simulations of the diffusivities (via dynamical Smagorinsky, slope limited diffusion or implicit large eddy simulation (ILES)).
Authors of these studies have been able to lower the diffusivity level low enough to yield more time dependent dynamo solutions possessing both cyclic behavior, regular butterfly diagram and torsional oscillation-like behavior
\citep{2011A&A...531A.162K,2012ApJ...755L..22K,2013ApJ...778...41K,2013ApJ...775...69D,2013JCoPh.236..608S,2013SoPh..282..335B,2013ApJ...777..153A}. Some have even obtained for the first time
buoyant magnetic wreaths \citep{2011ApJ...739L..38N,2013ApJ...762...73N}. In these recent simulations, magnetic wreath-like structures become turbulent and intermittent enough, that many intense bundles of fields reaching 50 kG start to form $\Omega$-loop like structures with statistical properties for rising loops that qualitatively agree with observations \citep{2014SoPh..289..441N}. We believe that such simulations are the progenitors of future more realistic {\it spot-dynamos} that will allow us to characterize the link between dynamo action, flux emergence and large scale magnetic topology as a function of stellar parameters.

Finally also note that the torque applied by the magnetized wind (see discussion in \S 1.5 and 2.1) depends mostly on the low order fields and that the field topology is expected to change as a function of stellar cycle and overall rotation rate \citep[see][]{2011ApJ...737...72P,reville14}. Hence assessing the large-scale magnetic topology of stellar magnetic fields and how it changes as a function of stellar parameters is very important as it has a direct bearing on stellar evolution.

\section{Future Directions}

The characterization of the many facets of stellar dynamics (convection, rotation, magnetism) has made tremendous progress over the last 20 years. We have moved away from a quasi-static view of stars to a time dependent, turbulent and magnetic description of these complex objects. Asteroseismology has significantly contributed to this progress, along with spectropolarimetry and ever improving numerical simulations and models of stellar convection and dynamo action. Still much remains to be done. Many of the assumptions have been deduced by extrapolating our (incomplete) understanding of our star, the Sun, to other stars. The continuous back and forth comparison between the Sun and other stars has been most productive and will continue being so with many projects relying both on ground-based telescopes and space-born missions in the planning.The launch of ESA's M3 Cosmic Vision PLATO in 2024 \citep{2013arXiv1310.0696R} will continue the rich tradition of data collection initiated by its predecessors Corot and \emph{Kepler}, helping refine our view of stellar dynamics and magnetism. PLATO has a field of view of more than 2200 square degrees and will observe two different fields for two years. It will combine huge statistics (about 300,000 stars) with high temporal resolution and length of observations to properly extract internal and surface stellar rotations as well as signatures of the existence of stellar cycles. Because one of the PLATO fields will include the \emph{Kepler} field, long-term magnetic variability (of the order of decades) will be potentially unveiled. Fields of
shorter observing times of six months (covering in total half of the sky) will add around a million new stars for which oscillations will be measured. This large amount of high-quality seismic data will be a treasure  trove for potential
new discoveries and will put new constraints on the theories of stellar activity. In the more immediate future, in 2017, the launch of NASA's new planet hunting mission TESS is planned. Although this mission is much smaller in scope than \emph{Kepler}, TESS will observe all stars in the sky down to magnitude ~12.5. Apart from the polar regions of the sky (observed for six months), other fields -- where the density of stars is higher -- will be observed for around 30 days. TESS will therefore contribute significantly to stellar rotation period studies, but it is by design focused on shorter timescale variability studies than PLATO.

Furthermore, other recently launched space missions have already or will soon provide asterosesimic measurements. Apart from the Canadian mission MOST, whose typical 3 weeks observations per target are not necessarily suited for the type of studies we have discussed here, the BRITE constellation of six nano-satellites (http://www.brite-constellation.at) and \emph{Kepler's} K2 missions are also contributing to the generation new constraints. The latter will be particularly important because it provides the ability to employ the new tools of asteroseismology in well-studied star clusters, such as M67, the Pleiades, and the Hyades.  K2 will also sample a variety of stellar populations in the Galaxy, important for our understanding of stellar population effects.

In addition to the space-borne instruments, the Danish-led ground-based Stellar Observations Network Group (SONG, \cite{2014IAUS..301...69G}) will soon provide very high-quality spectroscopic asterosesismic data of distant stars.
Asterosesismic data observed in velocity posses a much higher signal to noise ratio compared to the photomoetric data obtained in space, but for much less stars though. Two of the planned eight observing sites, distributed all over the Earth, are already funded, and either in construction (China) or already in operation (Teneriffe). The very high-quality data expected from SONG
will allow us to obtain, for example, information from the very outer stellar layers with unprecedented accuracy, information which is much needed to improve our knowledge about the still ill-understood surface-effects (see Section~1.2) and the details of the physical mechanisms that drive solar-like oscillations.

The detailed analysis of the legacy star catalog observed by \emph{Kepler}, the exciting and newly planned asteroseismic observation campaigns, the improvements in computation power and, last but not the least, our own improved understanding of the physical processes in stellar interior will not only bring a wealth of new discoveries, but will also bring stellar physics to a new level of sophistication.

On the theoretical side, recent advances based on analytical work and multi-D models of stars of various mass, rotation rates and age, have started paving the way to a direct comparison with asteroseismic and spectropolarimetric data.
Simulations coupling the angular evolution and magnetic activity on long time scale are in the making and should help explain the rotation history of stars. Acoustic and gravity modes excitation and propagation are now computed in detail and stellar cycles have now been found in many self-consistent convective dynamo simulations. General trends such as getting higher activity levels and shorter cycle period for faster rotating stars have been recovered. Still the exact dynamo mechanisms at work and resulting magnetic properties (topology, intensity, variability) for any given star are not yet fully elucidated. This comes about because the relative importance, spatial and temporal phasing and nonlinear interactions of the identified processes (e.g. shear in the tachocline, $\alpha$ and $\omega$ effects, stability of the magnetic wreaths, role
of the meridional flow and turbulent pumping) at the origin of the organisation of the global (large scale) dynamo and the establishment of a cyclic magnetic activity vary. Thus the characterisation of the exact dependence of the stellar dynamo properties to stellar parameters such as mass, rotation/age and stratification is still an evolving understanding. In order to progress further theoretical models and high performance multi-D numerical simulations will benefit from having the following
inputs/constraints:

\begin{itemize}
\item better constrained variations of the differential rotation in solar-like stars as a function of stellar parameters (mass, rotation/age, metallicity) and its latitudinal and radial
profiles
\item improved characterization of the $P_{cyc}$ vs $P_{rot}$ relation
\item some assesment of the surface horizontal flows
\item convective power and extent of the convective envelopes
\item link between star spot numbers and stellar parameters, relative contribution to stellar brightness between dark spot, and bright faculae
\item high cadence spectropolarimetric maps over the course of many years hence possibly mapping stellar cycles for many different stars
\item calibration with asteroseismology of gyrochronology by studying stellar clusters of various ages
\end{itemize}

It is our hope that in the next decade or so, these multitude of information will become available and will aid the community in constraining and further developing theories and computational models of solar and stellar activity.

\begin{acknowledgements}
This work was partly supported by ERC STARS2 207430 grant, ANR Toupies and IDEE grants, FP7 IRSES 269194 and SPACEINN 312844 programs, INSU/PNST. Funding for the Stellar Astrophysics Centre is provided by The Danish National Research Foundation (Grant agreement no.: DNRF106). The research is supported by the ASTERISK project (ASTERoseismic Investigations with SONG and \emph{Kepler}) funded by the European Research Council (Grant agreement no.: 267864). CESSI is supported by the Ministry of Human Resource Development, Government of India and IISER Kolkata. A.S.B and R.A.G. acknowledge support by the CNES via GOLF, CoRoT annd Solar Orbiter grants. D.N. acknowledges support from the Ramanujan Fellowship (Department of Science and Technology) and a grant from the (United States) Asian Office of Aerospace Research and Development. A.S. Brun is grateful to both the University of Kyoto, RIMS and Prof. M. Yamada, K. Shibata, H. Isobe \& S. Takahiro for their invitation in Fall 2013 and to the University of California at Santa Barbara, KITP and the organisers of the waves and flow program in Spring 2014 where most of this paper was written.
\end{acknowledgements}

\bibliographystyle{aps-nameyear}      
\bibliography{./Chapter13}



\end{document}